\documentclass[twocolumn,showpacs,nofootinbib,aps,10pt,prd]{revtex4}
\usepackage[T1]{fontenc}
\usepackage{graphicx,dcolumn,bm,amsfonts,amsmath,color,xcolor,amsthm,multirow}
\usepackage[colorlinks=true, citecolor=blue, linkcolor=blue, urlcolor=blue]{hyperref}
\usepackage{slashed}
\usepackage{ulem}
\usepackage{setspace}
\usepackage{txfonts}
\usepackage{cancel}
\DeclareMathAlphabet{\mathcal}{OMS}{cmsy}{m}{n}
\DeclareSymbolFont{largesymbols}{OMX}{cmex}{m}{n}

\allowdisplaybreaks
\begin{document}

\title{Scrutinizing lepton flavor universality and transition form factor correlation from charmed meson semileptonic decay into light strange vector $K^*$ meson}

\author{Sheng-Bo Wu$^*$}
\affiliation{Department of Physics, Guizhou Minzu University, Guiyang 550025, P.R.China}
\author{Dong Huang\footnote{Sheng-Bo Wu and Dong-Huang contributed equally to this work.}}

\author{Fang-Ping Peng}
\affiliation{Department of Physics, Guizhou Minzu University, Guiyang 550025, P.R.China}

\author{Hai-Bing Fu}
\affiliation{Department of Physics, Guizhou Minzu University, Guiyang 550025, P.R.China}

\author{Long-Zeng}
\email{zlong@cqu.edu.cn (corresponding author)}
\affiliation{Department of Physics, Guizhou Minzu University, Guiyang 550025, P.R.China}

\begin{abstract}
In this paper, we calculate the transition form factors (TFFs) $A_{1,2}(q^2)$ and $V(q^2)$ of $D_s^+ \to K^{*0}$ decays by using the QCD light-cone sum rule (QCD LCSR) and constructing a correlation function containing the usual current, in which the twist-2 transverse and longitudinal light-cone distribution amplitudes (LCDAs) $\phi^\lambda_{2;K^{*0}}(x,\mu_0)$ with $\lambda=(\bot,\|)$ of the $K^{*0}$ meson constitute the main source of theoretical uncertainty. Based on this, we construct these LCDAs using the light-cone harmonic oscillator model. Subsequently, the TFFs obtained in the large recoil region are $A_{1}(0)=0.579_{-0.028}^{+0.024}$, $A_{2}(0)=0.414_{-0.023}^{+0.021}$, and $V(0)=0.830_{-0.020}^{+0.020}$, and the corresponding ratios are obtained to be $r_V=1.433_{-0.090}^{+0.110}$ and $r_2=0.715_{-0.067}^{+0.075}$. Furthermore, we predict the correlation of TFFs and their corresponding ratios. Then, we extrapolate the TFFs over the whole physical $q^2$-region and calculate the decay widths and branching fractions for the semileptonic decays $D_s^+\to K^{*0}\ell^+\nu_{\ell}$. The results are $\mathcal{B}(D_s^+\to K^{*0}e^+\nu_{e})=(2.05_{-0.16}^{+0.13})\times 10^{-3}$ and $\mathcal{B}(D_s^+\to K^{*0}\mu^+\nu_{\mu})=(1.95_{-0.15}^{+0.13})\times 10^{-3}$. Meanwhile, we calculated the branching fraction ratio $\mathcal{R}_{\mu/e}^{K^{*0}}$ to be $0.950_{-0.002}^{+0.004}$. In addition, we extract values of the CKM matrix element, obtaining $|V_{cd}|_{(e\text{-Channel})} = 0.225_{-0.005}^{+0.005}$ and $|V_{cd}|_{(\mu\text{-Channel})} = 0.227_{-0.004}^{+0.011}$. Finally, we calculated the forward-backward asymmetry parameters for the $D_s^+\to K^{*0} \ell^+\nu_{\ell}$ decay.
\end{abstract}
\date{\today}
\maketitle
\newpage

\section{Introduction}
Semileptonic decays of $D_s^+$ mesons provide valuable information about the weak and strong interactions for hadrons composed of charm quarks~\cite{Antonelli:2009ws,Richman:1995wm}. The differential decay rate is directly proportional to the product of transition form factors (TFFs) and the Cabibbo-Kobayashi-Maskawa (CKM) matrix element $|V_{cd}|$~\cite{Kobayashi:1973fv}. The TFFs describe the strong interaction between initial and final state hadrons, while the CKM matrix element characterizes the mixing between different quark flavors in weak interactions. As a key component of the Standard Model (SM), this CKM matrix,  being a $3 \times 3$ unitary matrix governing transition probabilities between quark flavor states, must satisfy the exact unitarity condition. Taking the second row of this matrix as an example, one obtains $|V_{cd}|^2+|V_{cs}|^2+|V_{cb}|^2=1$. Although the contribution of $|V_{cd}|$ is relatively small, approximately $4.9\%$~\cite{ParticleDataGroup:2024cfk}, its precise determination remains of great physical interest. It is precisely the strong constraint imposed by CKM unitarity on $|V_{cd}|$ that renders it a precisely known input parameter. This makes semileptonic decays of $D_s^+$ mesons an ideal probe for accurately determining the TFFs and testing the self-consistency of SM.

In the experimental side, the semileptonic decay $D_s^+ \to K^{*0}\ell^+ \nu_\ell$, has attracted extensive experimental attention~\cite{CLEO:2009dyb,Hietala:2015jqa,BESIII:2018xre}. Among these experimental efforts, most measurements focus on the semileptonic decay $D_s^+ \to K^{*0}e^+ \nu_e$, while experimental investigations of the muonic channel $D_s^+ \to K^{*0} \mu^+ \nu_\mu$ remain scarce. For example, the CLEO Collaboration in 2009 and 2015 studied the semileptonic decay $D_s^+ \to K^{*0} e^+ \nu_e$ based on the $e^+e^-$ collision data samples collected by the CLEO-c detector at a center-of-mass energy of $4170\, \rm MeV$. The branching fractions obtained from the two measurements are $(1.80\pm0.70\pm0.10)\times 10^{-3}$ and $(1.80\pm0.40\pm0.10)\times 10^{-3}$, respectively~\cite{CLEO:2009dyb,Hietala:2015jqa}. Subsequently, the BESIII Collaboration performed a precise measurement of the semileptonic decay $D_s^+ \to K^{*0} e^+ \nu_e$ in 2019, using an $e^+e^-$ annihilation data sample with an integrated luminosity for $3.19\ \mathrm{fb}^{-1}$ collected at a center-of-mass energy of $4.178\, \rm GeV$~\cite{BESIII:2018xre}. The experimental Collaboration determined the TFFs ratios $r_V = V(0)/A_1(0) = 1.67 \pm 0.34_\text{stat} \pm 0.16_\text{syst}$ and $r_2 = A_2(0)/A_1(0) = 0.77 \pm 0.28_\text{stat} \pm 0.07_\text{syst}$ for the first time, and measured the decay branching fraction of $\mathcal{B}(D_s^+ \to K^{*0} e^+ \nu_e) = (2.37 \pm 0.26_\text{stat} \pm 0.20_\text{syst}) \times 10^{-3}$. Although the measurements for decay $D_s^+ \to K^{*0} e^+ \nu_e$ by BESIII Collaboration have been significantly improved compared with previous experiments, the corresponding muonic decay channel $D_s^+ \to K^{*0} \mu^+ \nu_\mu$ remains unmeasured. Therefore, further studies of this muonic channel will help comprehensively test lepton flavor universality (LFU). Therefore, further studies of this muonic channel will help comprehensively test LFU. In 2026, the BESIII Collaboration used a sample of $e^+e^-$ annihilation data with an integrated luminosity of $7.33\ \mathrm{fb}^{-1}$ collected at a center-of-mass energy of $4.128$  to $4.226\, \rm GeV$. To measure the semileptonic decay $D_s^+ \to K^*(892)^0 \mu^+ \nu_\mu$ for the first time, and also improved the precision for $D_s^+ \to K^*(892)^0 e^+\nu_e$ measurement~\cite{BESIII:2026ceu}. The measured branching fractions of the two decay channels are $(2.07 \pm 0.22_{\text{stat}} \pm 0.10_{\text{syst}}) \times 10^{-3}$ and $(2.14 \pm 0.18_{\text{stat}} \pm 0.10_{\text{syst}}) \times 10^{-3}$, respectively. Through the analysis of the dynamics for above two decay processes, the TFFs ratios of $D_s^+ \to K^*(892)^0$ transition were farther determined: $r_V = 1.63 \pm 0.14_{\text{stat}} \pm 0.08_{\text{syst}}$, $r_2 = 0.60 \pm 0.13_{\text{stat}} \pm 0.06_{\text{syst}}$, and $A_1(0) = 0.56 \pm 0.02_{\text{stat}} \pm 0.01_{\text{syst}}$, where the precision for the $r_V$ and $r_2$ was improved by a factor of one compared to previous results. Meanwhile, the collaboration measured for the first time the correlation between the TFF ratios $r_2$ and $r_V$ in the semileptonic decay $D_s^+ \to K^*(892)^0$, as well as the correlation between the branching fractions of $D_s^+ \to K^*(892)^0 e^+ \nu_e$ and $D_s^+ \to K^*(892)^0 \mu^+ \nu_\mu$, yielding correlation coefficients of $-0.28$ and $0.15$, respectively. In addition, this experimental study reports, for the first time in a model-independent way, the differential decay rates and lepton forward-backward asymmetries. Through the above results, the LFU test is carried out by using this decay in the whole $q^2$ region, finding no evidence for deviations from the SM predictions. These results describe for the first time the full dynamics of $D_s^+ \to K^*(892)^0$ transition and provide stringent experimental validation of various non-perturbative theoretical calculations.

From the theoretical perspective, calculations for semileptonic decays involving non-perturbative The TFFs should be performed using non-perturbative methods. In fact, many theoretical methods have predicted the semileptonic decay of $D_s^+ \to K^{*0}\ell^+\nu_{\ell}$, including the covariant confined quark model (CCQM)~\cite{Ivanov:2019nqd,Soni:2018adu}, chiral unitary approach ($\chi$UA)~\cite{Sekihara:2015iha}, light-front quark model (LFQM)~\cite{Cheng:2017pcq}, heavy meson chiral Lagrangians (HM$\chi$T)~\cite{Fajfer:2005ug}, constituent quark model (CQM)~\cite{Melikhov:2000yu}, light-cone sum rules (LCSRs)~\cite{Wu:2006rd,Lin:2025cmn}, large energy effective theory (LEET)~\cite{Palmer:2013yia}, covariant light-front quark model (CLFQM)~\cite{Zhang:2020dla,Verma:2011yw,Yang:2025gfz,Wang:2008ci}, relativistic quark model (RQM)~\cite{Faustov:2019mqr}, and holographic QCD (hQCD)~\cite{Ahmed:2023pod}. It is worth noting that although each method has its own limitations, combining them can provide a more comprehensive understanding and description of physical phenomena~\cite{Melikhov:2000yu}. For instance, theoretical predictions for the semileptonic decay $D_s^+\to K^{*0}\mu^+\nu_{\ell}$ within the CCQM framework, yielding the TFF ratios $r_2 = 0.99 \pm 0.20$, $r_V = 1.40 \pm 0.28$ and branching fraction $\mathcal{B}(D_s^+ \to K^{*0} \mu^+ \nu_\mu) = 1.7\,\times 10^{-3}$~\cite{Soni:2018adu}. Meanwhile, the covariant CLFQM can also be utilized to make theoretical predictions for semileptonic $D_s$ decays to vector mesons. Under this framework, the predicted TFF ratios and branching fractions are $r_V = 1.56$, $r_2 = 0.89$, and $\mathcal{B}(D_s^+ \to K^{*0}\mu^+\nu_\mu) = 1.76\times10^{-3}$~\cite{Yang:2025gfz}, respectively. Furthermore, the RQM based on the quasipotential approach has already been applied to $D_s$ decays into vector mesons. For this decay process, the TFF ratios $r_V = 1.61$ and $r_2 = 0.90$ were obtained, corresponding to branching fractions results of $2.10\,\times 10^{-3}$ and $2.00\,\times 10^{-3}$, respectively~\cite{Faustov:2019mqr}. By jointly analyzing the physical observables for semileptonic decay $D_s^+ \to K^{*0}\ell^+\nu_{\ell}$ as obtained from both experimental measurements and theoretical predictions, we find that discrepancies still exist between the current theoretical predictions and  experimental measurement. Therefore, to narrow the discrepancy between theoretical predictions and experimental measurements, further in-depth investigations of this decay process are for great physical significance, which constitutes the main motivation of the present work.

As a well-established theoretical framework, the LCSRs method incorporates both hard and soft contributions in the computation of hadron transitions. This enables it to provide useful insights into hadron physics problems spanning a wide range of energy scales. In this work, we investigate the $D_s^+ \to K^{*0}$ decay within this framework and calculation its TFFs by the usual current correlation function. Since the correlation function is constructed using usual currents, the resulting TFFs receive contributions from both the transverse and longitudinal twist-2 LCDAs for $K^{*0}$ meson. Physically, these TFFs essentially originate from the same hadronic matrix element and are parameterized by different Lorentz structures of the matrix element, so there is an intrinsic correlation between them~\cite{Li:2009tx}. This correlation is directly reflected in the fact that they share the same non-perturbative input LCDAs, so that the correlation between TFFs and their ratios can serve not only as a sensitive probe of the hadronic transition mechanism but also as a test of the self-consistency of theoretical calculations. Therefore, we adopt the light-cone harmonic oscillator (LCHO) model to construct the two twist-2 LCDAS for $K^{*0}$ meson, i.e., $\phi_{2;K^{*0}}^\bot(x,\mu)$ and $\phi_{2;K^{*0}}^\|(x,\mu)$. This model is based on the Brodsky-Huang-Lepage (BHL) framework~\cite{Brodsky:1981jv,Cao:1997hw,Huang:2004su}, which relates the equal-time wave function in the rest frame with the light-cone wave function (LCWF) in the light-cone frame through the Wigner-Melosh rotation effect. Subsequently, we derive the corresponding analytical expressions for these LCDAs. Meanwhile, this model has been widely applied in the research of LCDAs for various vector mesons~\cite{Hu:2024tmc, Fu:2016yzx, Yang:2025gcm, Choi:2007yu, Xu:2018mpf, Zhong:2023cyc, Wang:2025oix}.

The remaining parts of the paper are organized as follows: In Section~\ref{Sec:2}, we present the theoretical framework for studying the $D_s^+\to K^{*0}$, including the differential decay width, TFFs, and LCDAs. Section~\ref{Sec:3} is dedicated to the numerical analysis, where we provide the TFF, decay width, and branching fraction, etc. Section~\ref{Sec:4} contains a summary.

\section{Theoretical framework}\label{Sec:2}
In SM, the physical mechanism of the semileptonic decay $D_s^+\to K^{*0}\ell^+\nu_{\ell}$ can be described by the Feynman diagram shown in Fig.~\ref{Fig:FM}. During this decay process, the charm quark $c$ transitions into a strange quark $s$ through the exchange of a virtual $W^+$ boson. The virtual $W^+$ boson then transitions into the charged lepton $\ell^+$ and neutrino $\nu_{\ell}$. Among them, the anti-strange quark $s$ does not participate in the weak interaction at any stage of the process. For this semileptonic $c\to d\ell^+\nu_{\ell}$ transition, the effective Hamiltonian can be written as:
\begin{figure}[t]
\begin{center}
\centering
\includegraphics[width=0.31\textwidth]{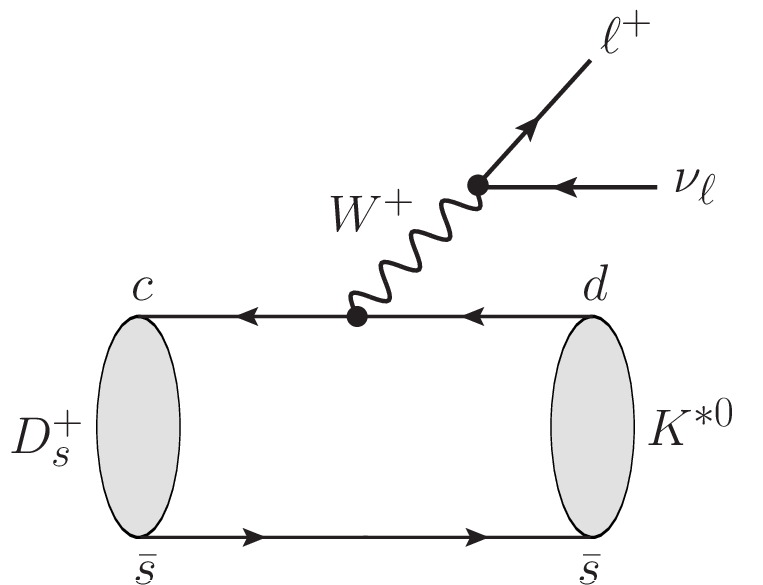}
\end{center}
\caption{The Feynman diagram for semileptonic decay $D_s^+ \to K^{*0}\ell^+\nu_\ell$, where the weak $c\to d\ell^+\nu_\ell$ transition is mediated by a virtual $W^+$ boson. Constituent quarks of the initial $D_s^+$ and final $K^{*0}$ mesons are explicitly drawn.}
\label{Fig:FM}
\end{figure}
\begin{align}
\mathcal{H}_{\text{eff}}^{\rm SM} =\frac{G_F}{\sqrt{2}} V_{cd} \, \big[ \bar{d} \gamma^\mu (1-\gamma_5) c \big] \, \big[ \bar{\ell} \gamma_\mu (1-\gamma_5) \nu_\ell \big],
\label{1}
\end{align}
where the Fermi coupling constant $G_F=1.166\times 10^{-5}\,\rm GeV^{-2}$ and $V_{cd}$ stands for the CKM matrix element. The $J^{\mu}=\bar{d} \gamma^\mu (1-\gamma_5) c$ and $L_{\mu}=\bar{\ell} \gamma_\mu (1-\gamma_5) \nu_\ell$ are expressed as quark and lepton current, respectively. Following the factorization theorem, the decay amplitude factorizes into the contraction for hadronic matrix element and leptonic currents as $\mathcal{M}(D_s^+\to K^{*0}\ell^+\nu_{\ell}) = \frac{G_F}{\sqrt{2}} V_{cd} H^\mu L_\mu$. Here the hadronic matrix element is $H^\mu = \langle K^{*0}(p,\lambda)|J^{\mu}|D_s^+(p+q)\rangle$, which encodes the non-perturbative strong interaction effects and can be fully parameterized by TFFs. Therefore, the precise calculation of the strong matrix element $H^\mu$ and lepton current $L_\mu$ is the key to determining the Invariant amplitude $\mathcal{M}(D_s^+\to K^{*0}\ell^+\nu_{\ell})$. From this, the differential decay width can be obtained.

The theoretical frameworks and calculation procedures for these two components are presented below. For the leptonic sector, a summation over the spins of the final-state leptons must be carried out when calculating the decay width. This requires the introduction of the leptonic tensor, which is defined as $L_{\mu \nu}=\mathrm{Tr}\left[ (\cancel{k}_1-m_{\ell})\gamma _{\nu}(1-\gamma _5)\cancel{k}_2\gamma _{\mu}(1-\gamma _5) \right]$. After taking the trace of this tensor using the Dirac equation, it can be written as
\begin{align}
L_{\mu\nu} = 8\left( k_1^{\mu} k_2^{\nu} + k_1^{\nu} k_2^{\mu} - (k_1\cdot k_2) g^{\mu\nu} - i \epsilon^{\mu\nu\alpha\beta} k_{1\alpha} k_{2\beta} \right),
\label{Eq:Lmu}
\end{align}
where the $k_1$ and $k_2$ represent the four momenta of the final-state charged lepton and neutrino, respectively. To match the helicity basis of the hadron part, we project the lepton tensor onto the helicity states of the virtual $W^+$ and define $L(m,n) = \epsilon^{\mu}(m)\epsilon^{*\nu}(n)L_{\mu\nu}$. Here, $m$ and $n$ label the helicity states, taking values $m,n=t,0,\pm$, corresponding to the time, longitudinal, and transverse polarization vectors, respectively. In the rest frame of $W^+$, its momentum and polarization vector take the following forms
\begin{align}
&q^\mu = (\sqrt{q^2}, 0, 0, 0),
\nonumber\\
&k_1^\mu = (E_1, |\mathbf{k}_1| \sin\theta_{\ell} \cos\chi, |\mathbf{k}_1| \sin\theta_{\ell} \sin\chi, |\mathbf{k}_1| \cos\theta),
\nonumber\\
&k_2^\mu = (|\mathbf{k}_1|,\!-|\mathbf{k}_1| \sin\theta_{\ell} \cos\chi,\!-|\mathbf{k}_1| \sin\theta_{\ell} \sin\chi, -\!|\mathbf{k}_1| \cos\theta_{\ell}),
\end{align}
where the $E_1 = (q^2 + m_\ell^2)/(2\sqrt{q^2})$ and $|\mathbf{k}_1| = (q^2 - m_\ell^2)/(2\sqrt{q^2})$ are the energy and charged lepton three-momentum in the $W^+$ rest frame. The polar angle $\theta_{\ell}$ is defined as the angle between the final vector meson momentum and the lepton $\ell $ momentum in the center-of-mass system. The azimuth $\chi$ is defined as the angle between the lepton decay plane and the vector meson decay plane. The polar angle $\theta$ and azimuthal angle $\chi$ span the ranges $0 \le \theta_{\ell} \le \pi$ and $0 \le \chi < 2\pi$, respectively. The helicity components of the polarization vector $\epsilon(\lambda_W)$ of the virtual $W^+$ boson in the present framework are given as
\begin{align}
\epsilon^\mu(0) &= (0, 0, 0, 1),\,
\,\epsilon^\mu(\pm)= \frac{(0, \mp 1, -i, 0)}{\sqrt{2}},
\nonumber\\
\epsilon^\mu(t)&= (1, 0, 0, 0).
\end{align}
in the above expression, the polarization vectors satisfy the orthogonality relation $\epsilon_\mu^*(m)\epsilon^\mu(n) = g_{mn}$ and completeness relation $\epsilon_\mu(m)\epsilon_\nu^*(n) = g_{\mu\nu}$. Here $g_{mn} = \operatorname{diag}(+, -, -, -)$. After integrating over the azimuth angle $\chi$, the matrix form for lepton tensor $L(m,n)$ can be arranged in the order of $t, +, 0, -$, and its explicit form is given as follows:
\begin{align}
(2q^2 v)^{-1}&L(m,n)(\theta)
\nonumber\\
&=
\begin{pmatrix}
0 & 0 & 0 & 0 \\
0 & (1 \mp \cos\theta_{\ell})^2 & 0 & 0 \\
0 & 0 & 2\sin^2\theta_{\ell} & 0 \\
0 & 0 & 0 & (1 \pm \cos\theta_{\ell})^2
\end{pmatrix}
\nonumber\\
&\,+ \delta_\ell
\begin{pmatrix}
4 & 0 & 4\cos\theta_{\ell} & 0 \\
0 & 2\sin^2\theta_{\ell} & 0 & 0 \\
4\cos\theta_{\ell} & 0 & 4\cos^2\theta_{\ell} & 0 \\
0 & 0 & 0 & 2\sin^2\theta_{\ell}
\end{pmatrix},
\label{matrixLmu}
\end{align}
where the velocity-type parameter $v = 1 - m_\ell^2/q^2$ and helicity-flip facyor $\delta_\ell = m_\ell^2/(2q^2)$.

Furthermore, in order to calculation the free-quark amplitude $\mathcal{M}(D_s^+ \to K^{*0}\ell^+\nu_\ell)$, Eq.~\eqref{1} must be sandwiched between the initial and final meson states, which gives the matrix element $\langle K^{*0}(p, \lambda) | J^{\mu}| D_s^+ (p + q) \rangle$. Since this matrix element is difficult to compute directly, it can be parametrized in terms of Lorentz-invariant TFFs $A_1(q^2)$, $A_2(q^2)$, $V(q^2)$ and $A_3(q^2)-A_0(q^2)$ as follows~\cite{Li:2009tx}:
\begin{align}
&\langle K^*(p,\lambda)|J^{\mu}|D_s^+(p+q)\rangle=-ie_\mu^{*(\lambda)}(m_{D_s^+}+m_{K^{*0}})A_1(q^2)
\nonumber\\
& \,\,\,+i(e^{*(\lambda)}\cdot q)\frac{(2p+q)_\mu}{m_{D_s^+}+m_{K^{*0}}}A_2(q^2)+iq_\mu(e^{*(\lambda)}\cdot q)\frac{2m_{K^{*0}}}{q^2}
\nonumber\\
& \,\,\,\times [A_3(q^2)-A_0(q^2)] + \epsilon_{\mu\nu\alpha\beta}e^{*(\lambda)\nu}q^\alpha p^\beta\frac{2V(q^2)}{m_{D_s^+}+m_{K^{*0}}}
\label{Eq:FF}
\end{align}
where $m_{D_s^+}$, $m_{K^{*0}}$ stand for the masses of $D_s^+$ and $K^{*0}$ meson, respectively. $e^{*( \lambda)}$ refers to the polarization vector for $K^{*0}$ meson with $\lambda=(\bot,\|)$. $p$ stands for the momentum of $K^{*0}$ meson, and $q$ denotes the momentum transfer. The TFFs $A_{1,2}(q^2)$ and $V(q^2)$ are associated with particle-exchange processes and correspond to exchange channels with quantum numbers $J^P=1^+$ and $J^P=1^-$~\cite{Richman:1995wm,Becirevic:2020rzi}, respectively. In addition, the TFF $A_3(q^2)$ is not independent and can be expressed in terms of $A_1(q^2)$ and $A_2(q^2)$ as
$A_3(q^2)=( m_{D_{s}^{+}}+m_{K^{*0}})/(2m_{K^{*0}}) A_1(q^2)-(m_{D_{s}^{+}}-m_{K^{*0}})/(2m_{K^{*0}}) A_2(q^2)
$. It also obeys the relation $A_3(0)=A_0(0)$.

To calculate the differential decay width for process $D_s^+ \to K^{*0} \ell^+ \nu_\ell$, we adopt the helicity formalism and parameterize the TFFs via the helicity amplitudes $H_{\pm,0,t}(q^2)$. The corresponding theoretical derivations are presented in the subsequent text. In the rest frame for $D_s^+$ , the polarization vectors $\epsilon^{*\mu}(\lambda_W)$ with $\lambda_W=(t,0,\pm)$ for virtual $W^+$ boson and $e^{*\mu}(\lambda_v)$ with $\lambda_v=(0,\pm)$ of $K^{*0}$ meson take the following forms.
\begin{align}
&\epsilon ^{*\mu}(t) =\frac{q^{\mu}}{\sqrt{q^2}},\,   \epsilon ^{*\mu}(0)=\frac{( q_0,0,0,|\mathbf{p}|)}{\sqrt{q^2}},   \epsilon ^{*\mu}(\pm)=\frac{(0,\mp 1,-i,0)}{\sqrt{2}}
\nonumber\\
&e^{*\mu}(\pm)=\frac{(0,\pm 1,-i,0)}{\sqrt{2}}, \,\,  e^{*\mu}(0)=\frac{(|\mathbf{p}|,0,0,-E_2)}{m_{K^{*0}}}
\end{align}
where the $|\mathbf{p}|=\lambda ^{1/2}(m_{D_{s}^{+}}^{2},m_{K^{*0}}^{2},q^2)/{2m_{D_{s}^{+}}^{2}}$ stand for three-momentum of $K^{*0}$ meson in the rest frame of $D_s^+$. The $E=(m_{D_{s}^{+}}^{2}+m_{K^{*0}}^{2}-q^2)/{2m_{D_{s}^{+}}^{2}}$ and $q_0=(m_{D_{s}^{+}}^{2}-m_{K^{*0}}^{2}+q^2)/{2m_{D_{s}^{+}}^{2}}$ stand for the momentum and energy of $K^{*0}$ meson in this frame, respectively.

According to Eq.~\eqref{Eq:FF}, the hadronic matrix element can be written as $H^{\mu} =e^{*\alpha}T_{\mu\alpha}$, in which $T^{\mu\alpha}$ represents a tensor containing TFFs. For the $D_s^+ \to K^{*0}$ transition within the helicity formalism, we contract the hadronic matrix element with the polarization vector $\epsilon^{*\mu}(m)$ of virtual $W^+$ boson and polarization vector $e^{*\alpha}(r)$ for the final vector meson. This permits the definition of helicity amplitudes: $H_{mr} = \epsilon^{*\mu}(m) e^{*\alpha}(r) T_{\mu\alpha}$. Therefore, the helicity component of hadron tensor $H(m,n)$ can be decomposed into the outer product of these helicity amplitudes: $H(m,n) = H_{mr}H_{nr}^{\dagger}$~\cite{Ivanov:2019nqd,Faustov:2019mqr}. The specific form is as follows,
\begin{align}
H(m,n)
&= \epsilon^{*\mu}(m)\epsilon^{\nu}(n)H_{\mu\nu}
\nonumber\\
&= \epsilon^{*\mu}(m)\epsilon^{\nu}(n)T_{\mu\alpha} e^{*\alpha}(r) e^{\beta}(s)\delta_{rs} T_{\beta\nu}^{\dagger} \nonumber\\
&= \epsilon^{*\mu}(m)e^{*\alpha}(r)T_{\mu\alpha} \cdot \left[\epsilon^{*\nu}(n)e^{*\beta}(s)T_{\nu\beta}\right]^{\dagger} \delta_{rs}
\nonumber\\
&\equiv H_{mr} H_{nr}^{\dagger}.
\label{Eq:Hmn}
\end{align}
where the $r$ and $s$ represents the helicity states for final vector meson, taking values $r, s = 0, \pm$. The $\delta_{rs}$ is Kronecker delta symbol, which reflects angular momentum conservation and relates the helicities of the vector meson and the virtual $W$ boson. From angular momentum conservation, we have $r = m $, $s = n$ for $m, n = \pm, 0$, and $r = s = 0$ for $m, n= t$. The non-zero helicity amplitudes are then obtained as follows~\cite{Ivanov:2019nqd}:
\begin{align}
H_{\pm}(q^2)&=\frac{\lambda ^{1/2}(m_{D_{s}^{+}}^{2},m_{K^{*0}}^{2},q^2)}{m_{D_{s}^{+}}+m_{K^{*0}}}\bigg[ \frac{(m_{D_{s}^{+}}+m_{K^{*0}})^2}{\lambda ^{1/2}(m_{D_{s}^{+}}^{2},m_{K^{*0}}^{2},q^2)}A_1(q^2)
\nonumber\\
&\mp V(q^2)\bigg] ,
\nonumber\\
H_0(q^2)&=\frac{1}{2m_{K^{*0}}\sqrt{q^2}}\bigg[ (m_{D_{s}^{+}}+m_{K^{*0}})(m_{D_{s}^{+}}^{2}-m_{K^{*0}}^{2}-q^2)A_1(q^2)
\nonumber\\
\,\,        & -\frac{\lambda (m_{D_{s}^{+}}^{2},m_{K^{*0}}^{2},q^2)}{m_{D_{s}^{+}}+m_{K^{*0}}}A_2(q^2) \bigg] ,
\nonumber\\
H_t(q^2)&=\frac{\lambda ^{1/2}(m_{D_{s}^{+}}^{2},m_{K^{*0}}^{2},q^2)}{\sqrt{q^2}}A_0(q^2).
\end{align}
where the $\lambda(m_{D_{s}^{+}}^{2},m_{K^{*0}}^{2},q^2)=(m_{D_s^+}^2+m_{K^{*0}}^2-q^2)^2-4m_{D_s^+}^2m_{K^{*0}}^2$ is K\"all\'en function. From the above formula, it can be seen that $A_0(q^2)$, $A_2(q^2)$, and $V(q^2)$ only contribute to $H_t(q^2)$, $H_0(q^2)$, and $H_{\pm}(q^2)$, respectively, while $A_1(q^2)$ contributes to both $H_0(q^2)$ and $H_{\pm}(q^2)$, and in the high-$q^2$ region, its contribution dominates the entire decay process.

Then, by contracting the helicity components of the hadron tensor $H(m,n)$, lepton tensor $L(m,n)$ and using the decomposition in Eq.~\eqref{Eq:Hmn}, we obtain the tensor contraction expression as follows:
\begin{align}
L^{\mu\nu}H_{\mu\nu}&=\!\sum_{m,n}L(m,n)H(m,n)= \!\sum_{m,n,r,s}L(m,n) H_{mr}H_{ns}^{\dagger}\delta_{rs}.
\label{eq:LHmn}
\end{align}
Subsequently, based on the result of Eq.~\eqref{eq:LHmn}, we take the squared modulus for invariant amplitude $\mathcal{M}$ and perform spin summation over all final-state particles. By further integrating over the full phase space, the differential decay width for semileptonic decay $D_s^+ \to K^{*0}\ell^+\nu_\ell$ is obtained within the helicity formalism, which depends on the momentum transfer $q^2$ and lepton polar angle $\cos\theta_{\ell}$. The explicit analytical expression is presented below.
\begin{align}
\!\frac{d\Gamma(D_{s}^{+}\!\to\!K^{*0}\ell ^+\nu _{\ell})}{dq^2d\cos \theta _{\ell}}&\!=\!\frac{G_{F}^{2}|V_{cd}|^2\lambda ^{1/2}(m_{D_{s}^{+}}^{2},m_{K^{*0}}^{2},q^2)(q^2-m_{\ell}^{2})^2}{512\pi^{3}m_{D_{s}^{+}}^{3}q^2}
\nonumber\\
&\times \big[(1+\cos^2 \theta _{\ell})\mathcal{H} _{\mathrm{U}}+2\big(1-\cos^2\theta_{\ell}\big)\mathcal{H}_{\mathrm{L}}
\nonumber\\
&+\,2\cos \theta_{\ell}\,\mathcal{H}_{\mathrm{P}}\,+\,\frac{m_{\ell}^{2}}{q^2}\,\big((1\,-\cos^2\theta_{\ell})\,\mathcal{H} _{\mathrm{U}}
\nonumber\\
&+2\cos ^2\theta _{\ell}\mathcal{H} _{\mathrm{L}}\,+\,2\mathcal{H} _{\mathrm{S}}-4\cos \theta _{\ell}\mathcal{H} _{\mathrm{SL}}\big)\big]
\label{Eq:dgamma0}
\end{align}
where the $m_{\ell}$ with  $\ell=(e,\mu)$ is lepton mass. The helicality structure function is defined as follows: $\mathcal{H}_\mathrm{U} =|H_+|^2+|H_-|^2$, $\mathcal{H}_\mathrm{L} =|H_0|^2$, $\mathcal{H}_\mathrm{P} =|H_+|^2-|H_-|^2$, $\mathcal{H}_\mathrm{SL} =\mathrm{Re}(H_0H_t^\dagger)$, and $\mathcal{H}_\mathrm{S}=|H_t|^2$~\cite{Hu:2024tmc}. Subsequently, integrating out $\cos\theta_{\ell}$ in Eq.~\eqref{Eq:dgamma0}, a differential decay width expression that only depends on $q^2$ can be obtained:
\begin{align}
&\frac{d\Gamma (D_{s}^{+}\!\to\! K^{*0}\ell ^+\nu _{\ell})}{dq^2}=\frac{G_{F}^{2}|V_{cd}|^2\lambda ^{1/2}(m_{D_{s}^{+}}^{2},m_{K^{*0}}^{2},q^2)( q^2-m_{\ell}^{2})}{192\pi ^3m_{D_{s}^{+}}^{3}q^2}
\nonumber\\
&\,\,\,\,\,\,\,         \times \bigg\{2\mathcal{A}^2A_{1}^{2}(q^2) + \frac{2\lambda (m_{D_{s}^{+}}^{2},m_{K^{*0}}^{2},q^2)}{\mathcal{A}^2}V^2(q^2 ) + \frac{1}{4m_{K^{*0}}^{2}q^2\mathcal{A}^2}
\nonumber\\
&\,\, \,\,\,\, \,\times          \bigg[\mathcal{A}^2\Delta A_1(q^2) -\lambda (m_{D_{s}^{+}}^{2},m_{K^{*0}}^{2},q^2)A_2(q^2)\bigg]^2\bigg( 1+\frac{m_{\ell}^{2}}{q^2}\bigg)
\nonumber\\
&\,\, \,\,\,\, \,         +\frac{3m_{\ell}^{2}}{2q^4}A_{0}^{2}( q^2)\bigg\},\label{Eq:dgamma1}
\end{align}
where we set $\mathcal{A}=(m_{D_s^+}+m_{K^{*0}})$ and $\Delta=(m_{D_s^+}^2-m_{K^{*0}}^2-q^2)$.  In Eq.~\eqref{Eq:dgamma1}, the TFFs $A_{0,1,2}(q^2)$ and $V(q^2)$ are crucial physical observable. Subsequently, we employ the QCD LCSR to derive the analytical expressions of the $D_s^+\to K^{*0}$ TFFs, starting from the following correlation function:
\begin{align}
\Pi_\mu(p,q)=i\int d^4xe^{iq\cdot x}\langle K^{*0}(p,\lambda) |T\{ j_{\mu}(x) ,j_{D_s^+}^{\dagger}(0)\} |0\rangle,
\label{correlators}
\end{align}
where the current $j_\mu(x) = \bar d(x)\gamma_\mu(1-\gamma_5)c(x)$ and $j_{D_s^+}^{\dagger}(0) = i m_c \bar s(0)\gamma_5 c(0)$. The matrix element from vacuum-to-$D_s^+$ meson $\langle D_s^+|\bar{c}im_c\gamma_5s|0\rangle=m_{D_s^+}^2f_{D_s^+}$, with the heavy meson mass $m_{D_s^+}$ and decay constant $f_{D_s^+}$. In which, the TFFs $A_{0,1,2}(q^2)$ and $V(q^2)$ enters the correlation function via the hadronic matrix element corresponding to the interpolating currents. They can be parameterized in terms of TFFs, as shown in Eq.~\eqref{Eq:FF}. Then, the invariant amplitude in the hadronic representation can be written as
\begin{align}
\Pi_i^{\rm H}&=\frac{m_{D_s^+}^2f_{D_s^+}(m_{D_s^+}+m_{K^{*0}})}{m_{D_s^+}^2-(p+q)^2}\tilde{A}_i(q^2)+\int_{s_0}^{\infty}\frac{\rho_i^{\rm H}}{s-(p+q)^2}ds
\nonumber\\
&+\cdots,\label{IA}
\end{align}
In the above equation, the symbol "$\cdots$" stands for the invariant amplitudes corresponding to high resonance states and continuum states, where the reduced functions $\tilde{A}_i$ with$i=(1,2,3,4$) are
\begin{align}
&\tilde{A}_1(q^2)=A_1(q^2),\,\tilde{A}_2(q^2)=\frac{A_2(q^2)}{(m_{D_s^+}+m_{K^{*0}})^2},
\nonumber\\
&\tilde{A_3}(q^2)=\frac{2m_{K^{*0}}[A_3(q^2)-A_0(q^2)]}{q^2(m_{D_s^+}+m_{K^{*0}})},
\nonumber\\
&\tilde{A_4}(q^2)=\frac{2V(q^2)}{(m_{D_s^+}+m_{K^{*0}})^2}.
\end{align}
In Eq.~\eqref{IA}, the ground state $D_s^+$ meson contributions have been isolated. Meanwhile, by replacing the contributions from the high resonance and continuum states with the dispersion relations, it is transformed into an integral form of spectral density $\rho_{i}^{\rm H}$. The possible subtraction terms and finite $q^2$-polynomial are neglected, as they have no contribution to the final rum rules. The continuum threshold $s_0$ set near the squared mass of the first excited $D_s^+$ meson state. Based on the quark-hadron duality ansatz, an approximate estimation of the hadron spectral density $\rho_{i}^{\rm H}$ is performed, satisfying $\rho_{i}^{\rm{H}} = \rho_{i}^{\rm{QCD}}\theta(s_0-s)$. In the space-like region, we employ the operator product expansion (OPE) with the $c$-quark propagator from Ref.~\cite{Huang:1998gp}, expressing the nonlocal matrix elements in terms of LCDAs of various twists~\cite{Ball:2004rg,Ball:1998sk,Ball:2007zt}. Meanwhile, by matching the hadronic and OPE representations after performing a Borel transform in $(p+q)^2$. Following the above standard LCSR calculation process, the analytical expressions for the $D_s^+ \to K^{*0}$ TFFs can be obtained. Specifically, as follows:
\begin{align}
A_1(q^2)&=B_{A_1} \phi_{2;{K^{*0}}}^\bot(u) + C_{A_1} \psi_{3;{K^{*0}}}^\|(u) + D_{A_1} \phi_{3;K^{*0}}^ \bot \left( u \right)
\nonumber\\
&-\, E_{A_1} \phi_{4;{K^{*0}}}^\bot(u)\! - F_{A_1} C_{K^{*0}}(u)\! + G_{A_1} I_L(u)\! + H_{A_1}
\nonumber\\
& \times\, H_3(u) + I_{A_1} \widetilde {\Psi}_{4;{K^{*0}}}^\bot (\underline \alpha) - J_{A_1} \Psi _{4;{K^{*0}}}^\bot (\underline \alpha) - K_{A_1}
\nonumber\\
& \times\, \Phi_{4;{K^{*0}}}^{\bot(1)} (\underline \alpha) + L_{A_1} \Phi_{4;{K^{*0}}}^{\bot(2)} (\underline \alpha) +\, M_{A_1} \big(\widetilde {\Phi} _{3;{K^{*0}}}^\parallel (\underline \alpha)
\nonumber\\
&+\,12 \Phi_{3;{K^{*0}}}^\parallel (\underline \alpha)\big),
\nonumber\\
\label{Eq:A1}
\\
A_2(q^2)&=B_{A_2} \phi_{2;{K^{*0}}}^\bot(u) - C_{A_2}\, \psi_{3;{K^{*0}}}^{\|}(u) - D_{A_2} \phi_{4;{K^{*0}}}^\bot(u)
\nonumber\\
&+\, E_{A_2} A_{K^{*0}}(u)\! - \!F_{A_2} B_{K^{*0}}(u)\! + \!G_{A_2} C_{K^{*0}}(u)\! + H_{A_2}
\nonumber\\
& \times \, I_L(u)\! - \!I_{A_2} H_3(u)\! + \!J_{A_2} \widetilde {\Psi}_{4;{K^{*0}}}^\bot (\underline \alpha)\! + \!K_{A_2} \Psi_{4;{K^{*0}}}^\bot (\underline \alpha)
\nonumber\\
& + \, L_{A_2} \Phi_{4;{K^{*0}}}^{\bot(1)} (\underline \alpha) + M_{A_2} \Phi_{4;{K^{*0}}}^{\bot(2)} (\underline \alpha),
\nonumber\\
\label{Eq:A2}
\\
V(q^2)&=B_{V} \phi_{2;{K^{*0}}}^\bot(u)\, + C_{V} \psi _{3;K^{*0}}^ \bot (u)\, - D_{V} \phi_{4;{K^{*0}}}^\bot (u)
\nonumber\\
& +\, E_{V} \widetilde {\Psi}_{4;{K^{*0}}}^\bot (\underline \alpha)\! + \!F_{V} \Psi_{4;{K^{*0}}}^\bot (\underline \alpha)\! - \!G_{V} \big(\Phi _{4;{K^{*0}}}^{\bot(1)} (\underline \alpha)
\nonumber\\
& -\, \Phi_{4;{K^{*0}}}^{\bot(2)} (\underline \alpha) \big),
\nonumber\\
\label{Eq:V}
\\
A_3(q^2)&-A_0(q^2)=B_{A_{30}} \phi_{2;K^{*0}}^\bot(u) - C_{A_{3-0}} \psi_{3;K^{*0}}^\|(u)
\nonumber\\
&+ D_{A_{30}}\phi_{4;K^{*0}}^\bot(u) - E_{A_{30}} A_{K^{*0}}(u) + F_{A_{30}} B_{K^{*0}}(u)
\nonumber\\
& +\, G_{A_{30}} C_{K^{*0}}(u) + H_{A_{30}} I_L(u) - I_{A_{30}} H_3(u)
\nonumber\\
& -\, J_{A_{30}} \Psi_{4;{K^{*0}}}^\bot (\underline \alpha) - K_{A_{30}} \Phi_{4;{K^{*0}}}^{\bot(1)} (\underline \alpha) - L_{A_{30}}
\nonumber\\
& \times\, \Phi_{4;{K^{*0}}}^{\bot(2)} (\underline \alpha) - M_{A_{30}} \widetilde {\Psi}_{4;{K^{*0}}}^\bot (\underline \alpha),
\label{Eq:A3-A0}
\end{align}
where $\underline \alpha=\{\alpha _1,\alpha _2,\alpha _3\}$ stand for the momentum fraction of three-particle quarks, antiquarks and gluons. The full explicit expressions for the auxiliary coefficients $B_{A_1}$, $C_{A_1}$, $D_{A_1}$, $E_{A_1}$, $F_{A_1}$, $G_{A_1}$, $H_{A_1}$, $I_{A_1}$, $J_{A_1}$, $K_{A_1}$, $L_{A_1}$, and $M_{A_1}$, etc., and others appearing in Eq.~\eqref{Eq:A1}-\eqref{Eq:A3-A0} are summarized in appendix~\ref{sec:appendixA} for easy reference. From the above expressions, the main theoretical uncertainties of TFFs $A_{0,1,2}(q^2)$ and $V(q^2)$ all arise from the twist DAs of the $K^{*0}$ meson. Among these, the twist-2 part is the most significant contributor and the dominant source of the current theoretical error.

For the twist-2 LCDAs of $K^{*0}$ meson, the current correlation function (\ref{correlators}) shows that these amplitudes dominate the $D_s^+ \to K^{*0}$ TFFs. To obtain the behavior of twist-2 LCDAs for the $K^{*0}$ meson, we construct the corresponding LCDAs within the LCHO model following the BHL prescription~\cite{Brodsky:1981jv, Cao:1997hw, Huang:2004su}. In general, the twist-2 LCDAs $\phi_{2;K^{*}}^{\lambda}(x,\mu_0)$ for $K^{*0}$ meson can be derived from its LCWF $\psi_{2;K^{*}}^{\lambda}(x,\mathbf{k}_\perp)$, and the defining relations between them are given as follows:
\begin{align}
\phi_{2;K^{*}}^{\lambda}(x,\mu_0) = \frac{2\sqrt{3}}{\tilde{f}_{K^{*}}^{\lambda}} \int_{|\mathbf{k}_\perp|^2 \le \mu_0^2} \frac{d^2\mathbf{k}_\perp}{16\pi^3} \, \psi_{2;K^{*}}^{\lambda}(x,\mathbf{k}_\perp),
\label{phi1}
\end{align}
where the $\lambda=(\|,\bot)$ corresponds to the longitudinal and transverse components of $\phi_{2;K^{*0}}^{\lambda}(x,\mu)$, respectively. The $\widetilde f_{K^{*0}}^\bot = f_{K^{*0}}^\bot / \sqrt{3}$ and $\widetilde f_{K^{*0}}^\| = f_{K^{*0}}^\| / \sqrt{5}$~\cite{Ball:2007zt}. In the LCHO model, the LCWF for the $K^{*0}$ meson can naturally be decomposed into two independent parts: the radial part and the spin-space part. Specifically, the radial part describes the relative momentum distribution of quark-antiquark pairs, typically described by the wave function from BHL prescription. The spin-space part, on the other hand, encodes the spin structure and the quark spin-orbital angular momentum coupling, and is closely related to the meson's polarization and helicity structure. Then, based on the BHL prescription, the LCHO model of $K^{*0}$ meson twist-2 LCWF can be given~\cite{Wu:2010zc,Wu:2011gf}:
\begin{align}
\psi_{2;K^{*0}}^{\lambda}(x,\mathbf{k}_\perp) = \sum_{h_1h_2} \chi_{h_1h_2}^{\lambda}(x,\mathbf{k}_\perp) \, \psi_{2;K^{*0}}^{R}(x,\mathbf{k}_\perp),
\label{2}
\end{align}
where $h_1$ and $h_2$ are the helicities of the two constituent quarks in $K^{*0}$ meson, respectively. The $\chi_{h_1h_2}^{\lambda}(x,\mathbf{k}_\perp)$ is their spin-space wave function, and $\psi_{2;K^{*0}}^R(x, \mathbf{k}_\perp)$ is the radial wave function. This spin-space wave function is obtained by the Wigner-Melosh rotation~\cite{Huang:1994dy,Huang:2004su,Wu:2005kq}, and its explicit form is given as follows.
\begin{align}
\chi_{h_1h_2}^{\lambda}(x,\mathbf{k}_\perp) = \frac{\bar{x}m_s + x m_q}{\sqrt{\mathbf{k}_\perp^2 + (\bar{x}m_s + x m_q)^2}} = \frac{\rm Y}{\sqrt{\mathbf{k}_\perp^2 + \rm Y^2}},
\label{3}
\end{align}
where $x$, $\bar{x}=1-x$ represent the momentum fractions of strange quarks $s$ and light quarks $q$, respectively. The $m_s$ and $m_q$ represent the constituent masses of the strange quark and light quark, respectively, with ${\rm{Y}} = \bar{x}m_s + x m_q$. Under this framework, the spatial wave function $\psi_{2;K^{*0}}^R(x, \mathbf{k}_\perp)$ can be further decomposed into two parts that depend on the transverse momentum $\mathbf{k}_\perp$ and momentum fraction $x$, respectively. Therefore, the spatial wave function for the $K^{*0}$ meson can be expressed as follows:
\begin{align}
\psi_{2;K^{*}}^{R}(x,\mathbf{k}_\perp) &\propto \left[1+B_{2;K^{*}}^{\lambda} C_1^{3/2}(\xi)\right]
\nonumber\\
& \,\times\exp\left[-b_{2;K^{*}}^{\lambda 2}\left(\frac{\mathbf{k}_\perp^2 + m_s^2}{x} + \frac{\mathbf{k}_\perp^2 + m_q^2}{\bar{x}}\right)\right],
\label{4}
\end{align}
where $\xi=2x-1$ and $C_1^{3/2}(\xi)$, the Gegenbauer polynomial. Subsequently, substituting Eqs.~\eqref{3}-\eqref{4} into Eq.~\eqref{2} and integrating over the transverse momentum $\mathbf{k}_\perp$, we obtain the analytical expression for the twist-2 LCDA of $K^{*0}$ meson. Its explicit form is given as follows:
\begin{align}
\phi _{2;K^{*0}}^{\lambda}&(x,\mu )=\frac{A_{2;K^{*0}}^{\lambda}\sqrt{3x\bar{x}}{\rm Y}}{8\pi^{3/2}\widetilde{f}_{K^{*0}}^{\lambda}b_{2;K^{*0}}^{\lambda}}\varphi_{2;K^{*0}}^\lambda(x)
\nonumber\\
&\!\times\exp\Bigg[-b_{2;K^{*0}}^{\lambda 2}\frac{\bar{x}\hat{m}_{s}^{2}\!+\!x\hat{m}_{q}^{2}\!-\!\mathrm{Y}^2}{x\bar{x}}\Bigg]
\nonumber\\
&\!\times\Bigg[\mathrm{Erf}\Bigg( b_{2;K^{*0}}^{\lambda}\sqrt{\frac{\mu^{2}\!+\!\mathrm{Y}^2}{x\bar{x}}}\Bigg) \!-\!\mathrm{Erf}\Bigg(b_{2;K^{*0}}^{\lambda}\sqrt{\frac{\mathrm{Y}^2}{x\bar{x}}}\Bigg)\Bigg],
\label{phi2}
\end{align}
where the constituent quark masses $\hat{m}_q = 300~\rm{MeV}$, $\hat{m}_s = 450~\rm{MeV}$~\cite{Wu:2007vi}. The error function is expressed as ${\rm Erf}(x) = 2\int_0^x e^{-t^2} dt/\sqrt{\pi}$. Furthermore, the three uncertain model parameters in Eq.~\eqref{phi2}, namely $A_{2;K^{*0}}^\lambda$, $B_{2;K^{*0}}^\lambda$, and $b_{2;K^{*0}}^\lambda$, can be further constrained by the following conditions, including the normalization condition for the twist-2 LCDA of $K^{*0}$ meson, i.e., $\int \phi_{2,K^{*0}}^\lambda(x) dx = 1$, average value of transverse momentum squared, which is $\langle{\bf k}_\bot^2 \rangle_{K^{*0}}^{1/2}=0.37(2){\,\rm GeV}$~\cite{Wu:2010zc,Huang:2013yya}, and Gegenbauer moment $a_n^{\lambda}$ of the twist-2 DA for $K^{*0}$ meson, where $a_1^\bot (1{\,\rm GeV}) = 0.04(3)$ and $a_1^\| (1{\,\rm GeV}) = 0.03(2)$~\cite{Ball:2007zt}. Moreover, for the $K^{*0}$ meson twist-3 LCDA, we adopt the Wandzura-Wilczek (WW) approximation~\cite{Ball:1997rj} to express it in terms of the twist-2 LCDA. The twist-4 contributions are treated following Ref.~\cite{Ball:2007zt}. Through the above constraints, we can determine the overall behavior of the $K^{*0}$ meson twist-2 LCDA. This result provides a solid theoretical basis for the subsequent theoretical calculation of physical observables.

\section{Numerical results and discussions}\label{Sec:3}
In the subsequent numerical analysis, we used the following parameters for the calculations. We take the $c$-quark mass $m_c=1.27(2)\,\rm GeV$, $K^{*0}$ meson mass $m_{K^{*0}} =0.892\,\rm GeV$, $D_s^+$ meson mass $m_{D_s^+}=1.9685\,\rm GeV$~\cite{ParticleDataGroup:2024cfk}. The $K^{*0}$ meson decay constants $f_{K^{*0}}^{\bot}=0.185(5)\,\rm GeV$ and $f_{K^{*0}}^{\parallel}=0.220(5)\,\rm GeV$~\cite{Ball:2007zt}. The $D_s^+$ meson decay constant $f_{D_s^+}=251.1\pm2.4\pm3.0\,\rm MeV$~\cite{BESIII:2021bdp}.

\begin{table}
\begin{center}
\renewcommand{\arraystretch}{1.5}
\setlength{\tabcolsep}{15pt}
\footnotesize
\caption{The numerical results of model parameters $A_{2;K^{*0}}^{\lambda}$, $B_{2;K^{*0}}^{\lambda}$, and $b_{2;K^{*0}}^{\lambda}$ for the $K^{*0}$ meson twist-2 LCDA $\phi^{\lambda}_{2;K^{*0}}(x,\mu)$ at the energy scales $\mu=1.0$ and $1.5\,{\rm GeV}$, respectively.}
\label {Tab:I}
\begin{tabular}{l l l l l}
\hline
$\mu$   &$A_{2;K^{*0}}^\bot$  &$B_{2;K^{*0}}^\bot$   &$b_{2;K^{*0}}^\bot$ \\
\hline
$1.0$              &36.0088              &$-\,$0.0563           &0.6844  \\
$1.5$              &34.1354              &$-\,$0.0593           &0.6846   \\
\hline
                   &$A_{2;K^{*0}}^\|$    &$B_{2;K^{*0}}^\|$     &$b_{2;K^{*0}}^\|$ \\
\hline
$1.0$              &33.2771              &$-\,$0.0669           &0.6849  \\
$1.5$              &31.5511              &$-\,$0.0702           &0.6851\\
\hline
\end{tabular}
\end{center}
\end{table}
\begin{figure}[t]
\begin{center}
\includegraphics[width=0.40\textwidth]{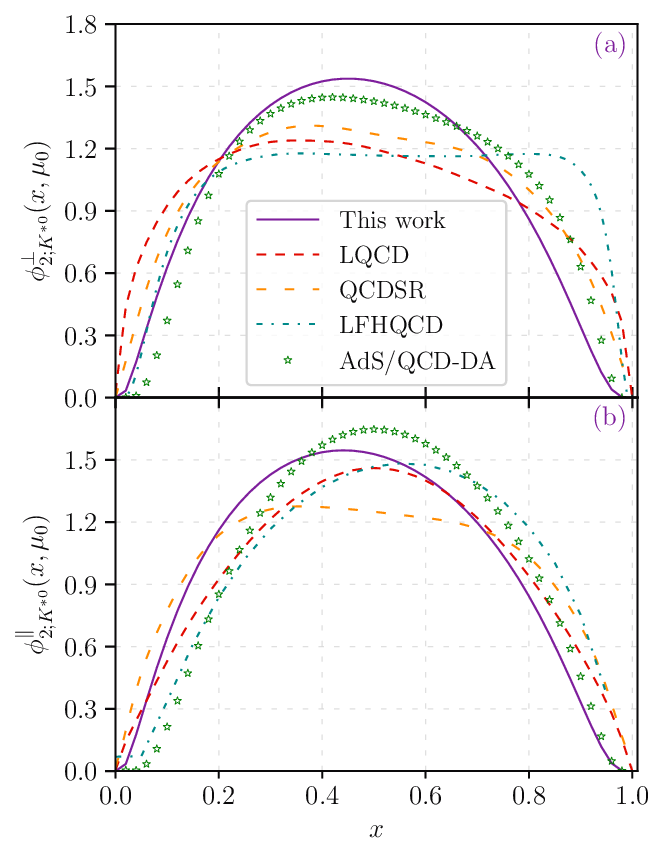}
\end{center}
\caption{ The twist-2 LCDA $\phi^{\lambda}_{2;K^{*0}}(x,\mu_0)$ with $\lambda=(\bot,\|)$ versus the momentum fraction $x$ at the scale $\mu_0 = 1\,\rm{GeV}$, where (a) and (b) correspond to the LCDA of transverse and longitudinal, respectively. For comparison, predictions from other theoretical models, including LQCD~\cite{Hua:2020gnw}, QCDSR~\cite{Ball:2007zt}, LFHQCD~\cite{Ahmady:2018fvo}, and AdS/QCD-DA~\cite{Ahmady:2013cva}, are also presented.}
\label{Fig:DA}
\end{figure}

Based on the above three constraints and Eq.~\eqref{phi2}, the LCHO model parameters of transverse and longitudinal twist-2 LCDA can be obtained, which are exhibited in Table~\ref{Tab:I}. In Table~\ref{Tab:I}, the values at $\mu=1.5\,{\rm GeV}$ will be used for subsequent calculation of the $D_s^+\to K^{*0}$ TFFs. Subsequently, we present the predicted results of $K^{*0}$ meson twist-2 LCDA $\phi^{\lambda}_{2;K^{*0}}(x, \mu_0)$ as a function of the momentum fraction $x$ under the scale $\mu_0=1\,{\rm GeV}$, as shown in Fig.~\ref{Fig:DA}. We also include predictions from LQCD~\cite{Hua:2020gnw}, QCDSR~\cite{Ball:2007zt}, holographic light-front QCD (LFHQCD)~\cite{Ahmady:2018fvo}, and AdS/QCD-DA~\cite{Ahmady:2013cva} for comparison. Our prediction prefers a single-peak behavior, while are consistent with the other theoretical predictions.
\begin{table}
\footnotesize
\renewcommand{\arraystretch}{1.4}
\setlength{\tabcolsep}{1.2pt}
\caption{The TFFs $A_{1,2}(0)$ and $V(0)$ at the large recoil of $D_s^+ \to K^{*0}$ transition. For comparison, we also present other theoretical and experimental research results.} \label {Tab:TFFs}
\begin{tabular}{llllll}
\hline
Method~~~~~~~~~~~~~~&$A_{1}(0)$~~~~~~~~~~~~~~~~~&$A_{2}(0)$~~~~~~~~~~~~~~~& $V(0)$~~\\
\hline
This work                               &$0.579_{-0.028}^{+0.024}$  &$0.414_{-0.023}^{+0.021}$  &$0.830_{-0.020}^{+0.020}$  \\
BESIII'26~\cite{BESIII:2026ceu}         &$0.56(2)(1)$       &/                          &/  \\
$\chi$UA'15~\cite{Sekihara:2015iha}     &$0.60$                     &$0.33$                     &$1.16$  \\
CQM'00~\cite{Melikhov:2000yu}           &$0.57$                     &$0.42$                     &$1.04$ \\
LCSR'06~\cite{Wu:2006rd}                &$0.589_{-0.042}^{+0.040}$  &$0.315_{-0.018}^{+0.024}$  &$0.771_{-0.049}^{+0.049}$ \\
LCSR'25~\cite{Lin:2025cmn}              &$0.42$                     &$0.21$                     &$0.82$  \\
LEET'13~\cite{Palmer:2013yia}           &$0.59$                     &$0.32$                     &$0.77$  \\
CLFQM~\cite{Zhang:2020dla,Verma:2011yw} &$0.56$                     &$0.46$                     &$0.87$  \\
CLFQM'26~\cite{Yang:2025gfz}            &$0.549_{-0.003}^{+0.003}$            &$0.485_{-0.002}^{+0.001}$  &$0.860_{-0.010}^{+0.010}$ \\
CLFQM'08~\cite{Wang:2008ci}             &$0.53$                     &$0.49$                     &$0.79$  \\
RQM'19~\cite{Faustov:2019mqr}           &$0.596$                    &$0.540$                    &$0.959$  \\
\hline
\end{tabular}
\end{table}

To calculate the TFFs $A_{1,2}(q^2)$ and $V(q^2)$ of $D_s^+ \to K^{*0}$ transition, it is also necessary to choose appropriate continuum threshold $s_0$ and Borel parameter $M^2$. These parameters can be determined according to the self-consistency criteria of QCDSR~\cite{Tian:2023vbh}. Based on this, the continuum threshold $s_0$ and Borel parameter $M^2$ corresponding to TFFs can be further obtained, and the results are $s_0^{A_1}=s_0^{A_2}=8.5\pm0.5\,\rm GeV^2$, $s_0^{V}=14\pm0.5\,\rm GeV^2$, $M_{A_1}^{2}=15.0\pm1.0\,\rm GeV^2$, $M_{A_2}^{2}=12.0\pm1.0\,\rm GeV^2$ and $M_{V}^{2}=15.0\pm1.0\,\rm GeV^2$, respectively. Then, we calculated TFFs at the large recoil region, and the results are presented in Table~\ref{Tab:TFFs}. For comparison, we have also presented the results derived from the latest experimental result and various theoretical approaches, including BESIII'26~\cite{BESIII:2026ceu}, $\chi$UA'15~\cite{Sekihara:2015iha}, CQM'00~\cite{Melikhov:2000yu},  LCSR~\cite{Wu:2006rd,Lin:2025cmn}, LEET'13~\cite{Palmer:2013yia}, CLFQM~\cite{Zhang:2020dla,Verma:2011yw,Wang:2008ci,Yang:2025gfz}, and RQM'19~\cite{Faustov:2019mqr}. The comparative analysis shows that within the range of uncertainty, our prediction results are in good agreement with the BESIII'26 measurements and theoretical predictions. However, the calculated results exhibit some deviation from the theoretical predictions for LCSR'25, which is mainly attributed to effects such as the continuum threshold $s_0$ and Borel parameter $M^2$ selecting different, higher-twist DA corrections.
\begin{table}
\footnotesize
\renewcommand{\arraystretch}{1.4}
\setlength{\tabcolsep}{5.0pt}
\caption{The TFF ratios $r_V$ and $r_2$ of $D_s^+\to K^{*0}$ TFFs. For comparison, we also presented experimental results and other theoretical predictions.} \label {Tab:RR}
\begin{tabular}{l l l}
\hline
Method~~~~~~~~~~~~~~~~~&$r_V$~~~~~~~~~~~~~~~~~~~~~~&$r_2$~~~~\\
\hline
This work                                &$1.433_{-0.090}^{+0.110}$    &$0.715_{-0.067}^{+0.075}$ \\
PDG~\cite{ParticleDataGroup:2024cfk}     &$1.70\pm0.4$               &$0.77\pm0.29$ \\
BESIII'19~\cite{BESIII:2018xre}          &$1.67\pm0.38$              &$0.77\pm0.29$ \\
BESIII'26~\cite{BESIII:2026ceu}          &$1.63\pm0.14\pm0.08$       &$0.60\pm0.13\pm0.06$ \\
CCQM~\cite{Ivanov:2019nqd,Soni:2018adu}  &$1.40\pm0.28$              &$0.99\pm0.20$ \\
LFQM'17~\cite{Cheng:2017pcq}             &$1.55$                     &$0.82$ \\
HM$\chi$T'05~\cite{Fajfer:2005ug}        &$1.93$                     &$0.55$ \\
CQM'00~\cite{Melikhov:2000yu}            &$1.82$                     &$0.74$ \\
LCSR'06~\cite{Wu:2006rd}                 &$1.31$                     &$0.53$ \\
LCSR'25~\cite{Lin:2025cmn}               &$1.95_{-0.16}^{+0.28}$     &$0.50_{-0.20}^{+0.13}$ \\
LEET'13~\cite{Palmer:2013yia}            &$1.53$                     &$0.91$ \\
CLFQM~\cite{Zhang:2020dla,Verma:2011yw}  &$1.55$                     &$0.82$ \\
CLFQM'26~\cite{Yang:2025gfz}             &$1.56$                     &$0.89$ \\
CLFQM'08~\cite{Wang:2008ci}              &$1.51$                     &$0.92$ \\
RQM'19~\cite{Faustov:2019mqr}            &$1.61$                     &$0.90$ \\
hQCD'23~\cite{Ahmed:2023pod}             &$1.53$                     &/ \\
\hline
\end{tabular}
\end{table}
\begin{figure*}[t]
\begin{center}
\centering
\includegraphics[width=0.40\textwidth]{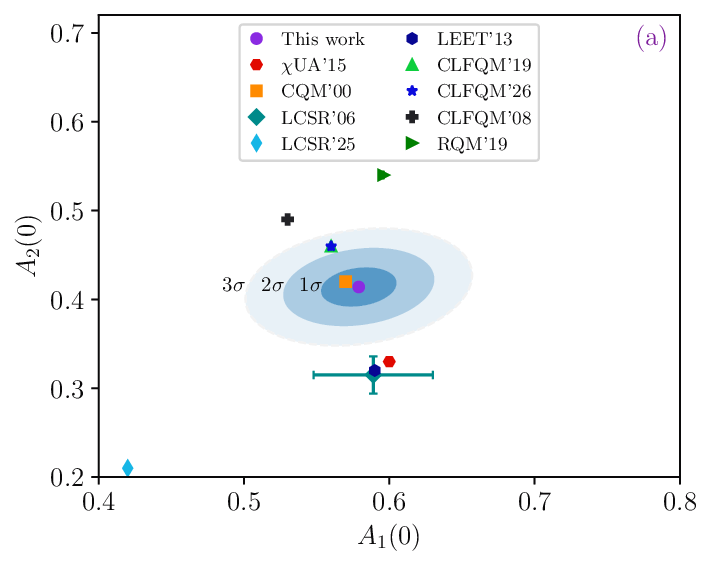}
\includegraphics[width=0.40\textwidth]{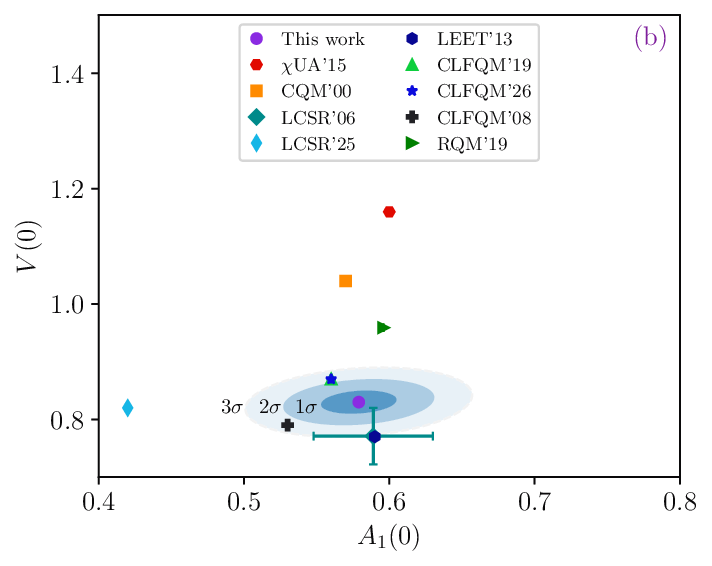}
\includegraphics[width=0.40\textwidth]{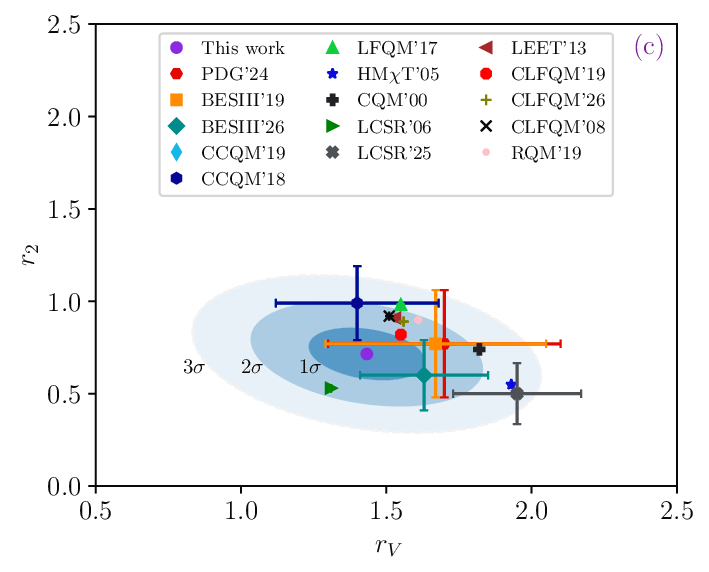}
\end{center}
\caption{Predictions for the correlation between the TFFs $A_2(0)$ with $A_1(0)$, $V(0)$ with $A_1(0)$, and the corresponding ratios $r_2$ and $r_V$ at the large recoil region of $D_s^+ \to K^{*0}$ transition, with various experimental measurements and nonperturbative theoretical calculations collected for comparison. The correlation coefficients are respectively set to $0.2$ and $-0.28$\,\cite{BESIII:2026ceu}. The nested shaded ellipses represent the $1\sigma$, $2\sigma$, and $3\sigma$ confidence regions of our theoretical results.}
\label{Fig:ratios}
\end{figure*}

Meanwhile, we also calculate the ratios of TFFs at the large recoil, and the results are presented in Table~\ref{Tab:RR}. From Table~\ref{Tab:RR}, it can be seen that the ratios $r_V$ and $r_2$ for the $D_s^+ \to K^{*0}$ TFFs obtained in this work are consistent with most experimental measurements and theoretical predictions within errors, demonstrating good self-consistency. Specifically, our result for $r_V$ is close to those from CCQM~\cite{Ivanov:2019nqd,Soni:2018adu}, LFQM'17~\cite{Cheng:2017pcq}, LEET'13~\cite{Palmer:2013yia}, CLFQM'08~\cite{Wang:2008ci}, and hQCD'23~\cite{Ahmed:2023pod}, but is slightly lower than the higher predictions from PDG~\cite{ParticleDataGroup:2024cfk} and BESIII~\cite{BESIII:2018xre,BESIII:2026ceu}, as well as CQM'00~\cite{Melikhov:2000yu}, HM$\chi$T'05~\cite{Fajfer:2005ug}, and LCSR'25~\cite{Lin:2025cmn}. For $r_2$, our result agrees well with those from BESIII'26~\cite{BESIII:2026ceu}, CQM'00~\cite{Melikhov:2000yu}, and HM$\chi$T'05~\cite{Fajfer:2005ug}, but is below those from CCQM~\cite{Ivanov:2019nqd,Soni:2018adu}, LEET'13~\cite{Palmer:2013yia}, CLFQM'08~\cite{Wang:2008ci}, and RQM'19~\cite{Faustov:2019mqr}. Although our prediction results differ from the above experimental and theoretical results, we can further evaluate their rationality by examining the overall trend of TFF. Furthermore, in order to deepen the understanding of TFF, we present in Fig.~\ref{Fig:ratios} the theoretical predictions for the correlations of $A_1(0)$, $A_2(0)$, $V(0)$, and their ratios $r_2$ and $r_V$, and compare them with other theoretical and experimental results. Since these TFFs describe the same strong-particle transition process via the same non-perturbative matrix elements, they are significantly correlated. This correlation directly affects decay amplitudes and observables and also serves as a test of the SM's self-consistency. Fig.~\ref{Fig:ratios}(a)-(b) show that the predictions of this paper for the correlations of $A_2(0)$ with $A_1(0)$ as well as $V(0)$ with $A_1(0)$ have a relatively large dispersion and uncertainty within the 1$\sigma$ to 3$\sigma$ range. The correlation of TFF ratios $r_V$ and $r_2$ is shown in Fig.~\ref{Fig:ratios}(c), and it is also compared with the latest measurements of BESIII, the average value of PDG, and various theoretical methods. The majority of theoretical results fall within the 2$\sigma$ ellipse and are consistent with this paper. $r_V$ and $r_2$, as strong particle observables that do not depend on lepton flavor, are crucial for testing the universality of lepton flavor. If future high-precision measurement results significantly deviate from the 3$\sigma$ ellipse, it may suggest the existence of new physics beyond the SM. This two-dimensional correlation comparison directly reveals dynamical information about strong-interaction TFFs and provides self-consistent input for physical observables. Therefore, further research on these physical quantities is a key direction for future experimental and theoretical work.
\begin{figure*}[t]
\begin{center}
\centering
\includegraphics[width=0.8\textwidth]{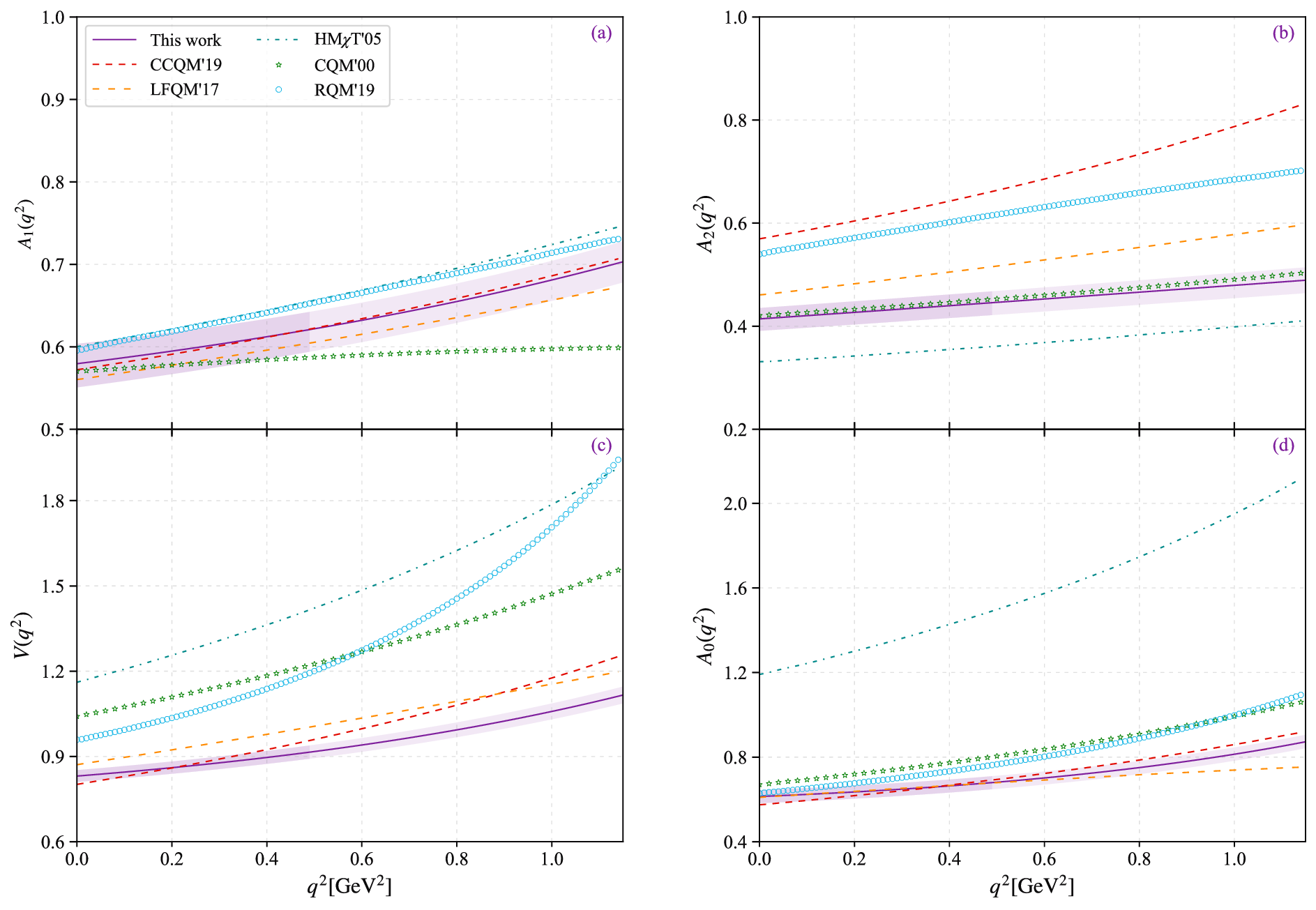}
\end{center}
\caption{The $D_s^+\to K^{*0}$ TFFs $A_{0,1,2}(q^2)$ and $V(q^2)$ versus $q^2$. For comparison, predictions from other theoretical models, including CCQM'19~\cite{Ivanov:2019nqd}, LFQM'17~\cite{Cheng:2017pcq}, HM$\chi$T'05~\cite{Fajfer:2005ug}, CQM'00~\cite{Melikhov:2000yu}, and RQM'19~\cite{Faustov:2019mqr}, are also presented.}
\label{Fig:fp2}
\end{figure*}

In order to obtain the behaviors of TFFs over the full $q^2$ region, the simplify series expansion (SSE) is adopted to extrapolated LCSR predictions into the full physically allowable region $0 \leq q^2 \leq (m_{D_s^+} - m_{K^{*0}})^2 \approx 1.15\,{\rm GeV}^2$. The reason is that the LCSR approach is only valid for low and intermediate $q^2$ region, and extrapolations to large $q^2$ region become unreliable. Specifically, the TFFs can be expanded as follows:
\begin{eqnarray}
F_i(q^2) =\frac{1}{1-q^2/m_{R,i}^2}\sum_{k=0,1,2}{\beta _{k,i}\,z^k( q^2,t_0 )},
\end{eqnarray}
where $F_i(q^2)$ with $i=(0,1,2,3,)$ represent $A_{0,1,2}(q^2)$ and $V(q^2)$, respectively, and $m_{R,i}$ stands for the masses of low-lying $D_s^+$ meson resonances, whose values are taken from the PDG~\cite{ParticleDataGroup:2024cfk}. The $\beta_{k,i}$ are real coefficients and $z(q^2,t)$ is the function,
\begin{eqnarray}
z^k( q^2,t_0 ) =\frac{\sqrt{t_+-q^2}-\sqrt{t_+-t_0}}{\sqrt{t_+-q^2}+\sqrt{t_+-t_0}},
\end{eqnarray}
in which $t_{\pm} = (m_{D_s^+} \pm m_{K^{*0}})^2$ and $t_0=t_{\pm}(1-\sqrt{1-t_-/t_+})$. The SSE method remains correct the analytic structure over the complex plane, while ensuring the expected scaling $F_i(q^2)\sim 1/q^2$. After analytically continuing the $D_s^+ \to K^{*0}$ TFFs to the full physical $q^2$ region, we then obtained the complete behaviors of TFFs $A_{0,1,2}(q^2)$ and $V(q^2)$, as showed in Fig.~\ref{Fig:fp2}. Meanwhile, we present the predictions of various theoretical approach such as CCQM'19~\cite{Ivanov:2019nqd}, LFQM'17~\cite{Cheng:2017pcq}, CQM'00~\cite{Melikhov:2000yu}, HM$\chi$T'05~\cite{Fajfer:2005ug}, and RQM'19~\cite{Faustov:2019mqr} for comparison. The analysis indicates that for TFF $A_1(q^2)$, the predictions in the low and intermediate $q^2$-regions are in good agreement with all reference theories, while over the entire $q^2$-range they are consistent with the results of CCQM'19~\cite{Ivanov:2019nqd} and LFQM'17~\cite{Cheng:2017pcq}. For $A_2(q^2)$, the predicted trend is generally consistent with the above models, and within uncertainties it agrees well with CQM'00~\cite{Melikhov:2000yu} over the full $q^2$-range. For $V(q^2)$, the low and intermediate $q^2$ predictions agree with CCQM'19~\cite{Ivanov:2019nqd} but are lower than those of other models. For $A_0(q^2)$, its behavior is consistent with CCQM'19~\cite{Ivanov:2019nqd}, LFQM'17~\cite{Cheng:2017pcq}, CQM'00~\cite{Melikhov:2000yu}, and RQM'19~\cite{Faustov:2019mqr}, but is lower than the prediction from HM$\chi$T'05~\cite{Fajfer:2005ug}. These differences mainly arise from the variations in input parameters and theoretical treatments among the different models.
\begin{figure}[t]
\begin{center}
\centering
\includegraphics[width=0.40\textwidth]{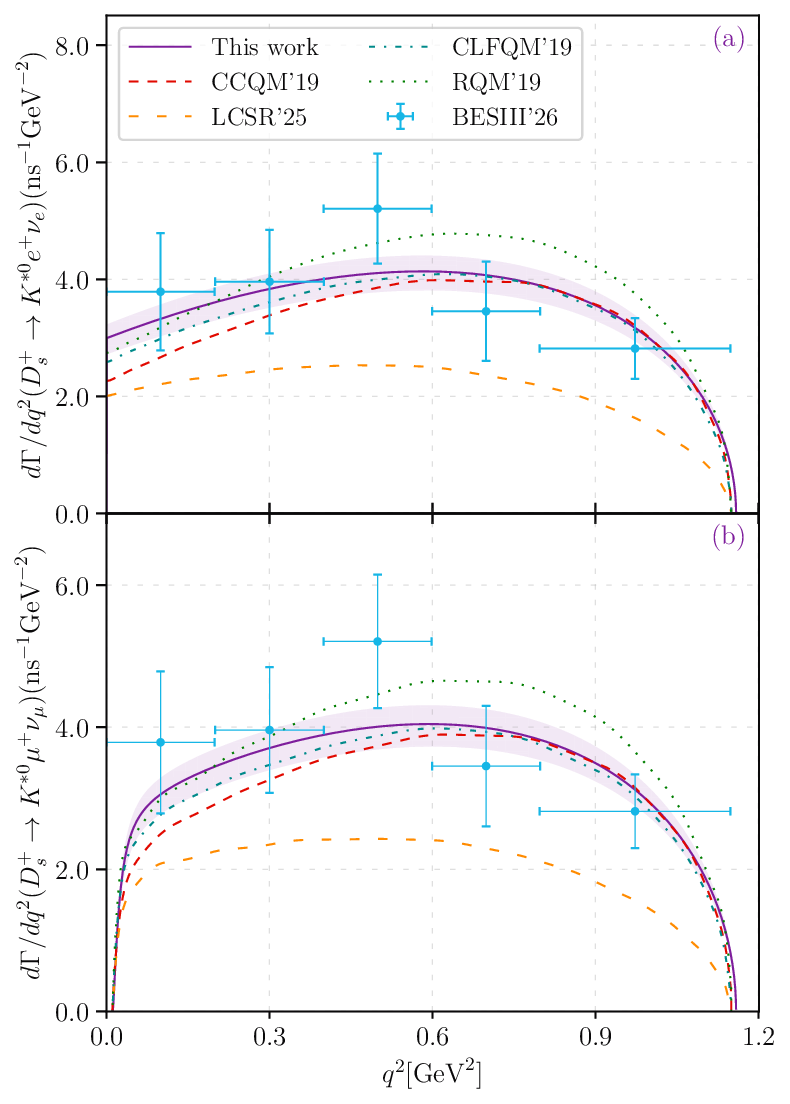}
\end{center}
\caption{The differential decay width for $d\Gamma/dq^2(D_s^+\to K^{*0}\ell^+\nu_{\ell})$ with $\ell=(e,\mu)$ versus $q^2$. Meanwhile, we also compare it with the experimental and theoretical results of BESIII'26~\cite{BESIII:2026ceu}, CCQM'19~\cite{Ivanov:2019nqd}, LCSR'25~\cite{Lin:2025cmn}, CLFQM'19~\cite{Zhang:2020dla}, and RQM'19~\cite{Faustov:2019mqr}.}
\label{Fig:dGamma}
\end{figure}
\begin{table}
\footnotesize
\renewcommand{\arraystretch}{1.0}
\setlength{\tabcolsep}{2.2pt}
\caption{Our predictions for the branching fractions of semileptonic decays $D_s^+ \to K^{*0}\ell^+\nu_{\ell}$ with $\ell=(e,\mu)$ are compared with results from other experimental and theoretical results (in unit $10^{-3}$).} \label {Tab:III}
\begin{tabular}{lllll}
\hline
Method~~~~~~~~~~~~~~~~~&$D_s^+\to K^{*0} e^+\nu_{e}$~~~~~~~~~~~~~&$D_s^+\to K^{*0} \mu^+\nu_{\mu}$~~~ \\
\hline
This work                                   &$2.05_{-0.16}^{+0.13}$                       &$1.95_{-0.15}^{+0.13}$ \\
PDG'24~\cite{ParticleDataGroup:2024cfk}     &$2.05\pm0.20$                                & / \\
CLEO'09~\cite{CLEO:2009dyb}                 &$1.80\pm0.70\pm0.10$                         & / \\
CLEO'15~\cite{Hietala:2015jqa}              &$1.80\pm0.40\pm0.10$                         & / \\
BESIII'19~\cite{BESIII:2018xre}             &$2.37\pm0.26\pm0.20$                         & / \\
BESIII'26~\cite{BESIII:2026ceu}             &$2.14\pm0.18\pm 0.10$                        & $2.07\pm0.22\pm 0.10$ \\
CCQM'19~\cite{Ivanov:2019nqd}               &$1.80$                                       & $1.80$ \\
CCQM'18~\cite{Soni:2018adu}                 &$1.80$                                       & $1.70$\\
$\chi$UA'15~\cite{Sekihara:2015iha}         &$2.02$                                       & $1.89$ \\
LFQM'17~\cite{Cheng:2017pcq}                &$1.90\pm0.20$                                & $1.90\pm0.20$ \\
HM$\chi$T'05~\cite{Fajfer:2005ug}           &$2.20$                                       & $2.20$ \\
CQM'00~\cite{Melikhov:2000yu}               &$1.90$                                       & $1.90$ \\
LCSR'06~\cite{Wu:2006rd}                    &$2.33_{-0.30}^{+0.29}$                       & $2.24_{-0.29}^{+0.27}$ \\
LCSR'25~\cite{Lin:2025cmn}                  &$1.55_{-0.20}^{+0.30}$                       & $1.48_{-0.19}^{+0.29}$ \\
CLFQM'19~\cite{Zhang:2020dla}               &$1.90$                                       & $1.82$\\
CLFQM'26~\cite{Yang:2025gfz}                &$1.85_{-0.01-0.07-0.03}^{+0.01+0.07+0.03}$   & $1.76_{-0.01-0.06-0.02}^{+0.01+0.06+0.03}$\\
CLFQM'08~\cite{Wang:2008ci}                 &$1.70$                                       & $1.70$\\
RQM'19~\cite{Faustov:2019mqr}               &$2.10$                                       & $2.00$\\
\hline
\end{tabular}
\end{table}
\begin{figure}[t]
\begin{center}
\centering
\includegraphics[width=0.40\textwidth]{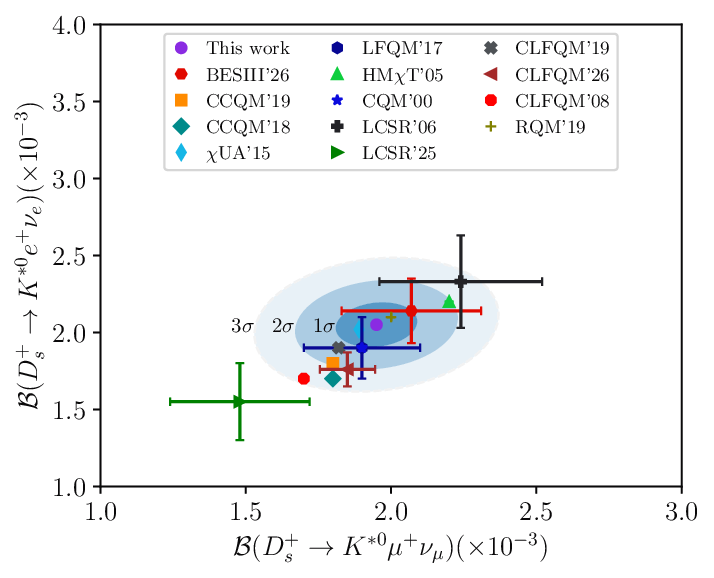}
\end{center}
\caption{We present predictions for the branching ratios of $D_s^+\to K^{*0}e^+\nu_e$ and $D_s^+ \to K^{*0}\mu^+\nu_\mu$, and compare them with available experimental and theoretical data. The correlation coefficient is taken as 0.15 from the most recent BESIII result~\cite{BESIII:2026ceu}. The nested shaded ellipses represent the $1\sigma$, $2\sigma$, and $3\sigma$ confidence regions of our theoretical results.}
\label{Fig:BRs}
\end{figure}
\begin{figure}[t]
\begin{center}
\centering
\includegraphics[width=0.40\textwidth]{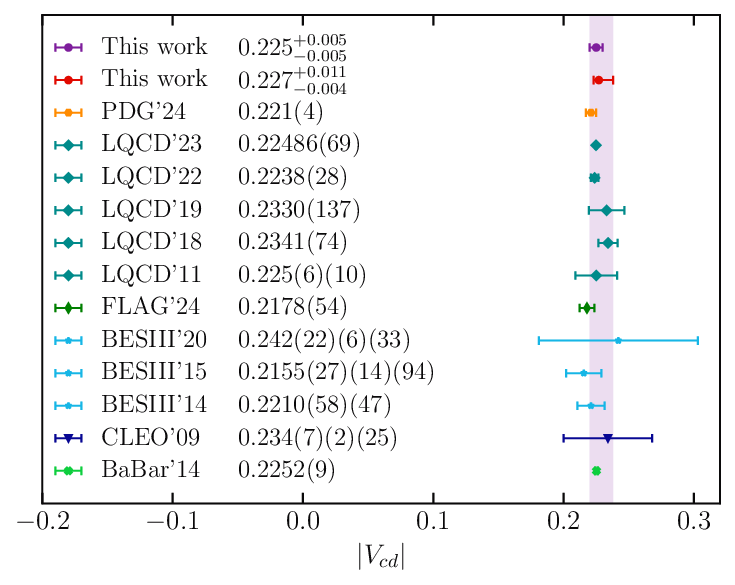}
\end{center}
\caption{Our prediction of $|V_{cd}|$ from $D_s^+\to K^{*0} e^+\nu_{e}$ and $D_s^+\to K^{*0} \mu^+\nu_{\mu}$ are presented. Furthermore, we also compare it with the results of other theoretical and experimental groups.}
\label{Fig:CKM}
\end{figure}

The differential decay widths is calculated using Eq.~\eqref{Eq:dgamma1} and the CKM matrix $|V_{cd}|$~\cite{ParticleDataGroup:2024cfk}. Specifically, we present the results for electron and muon decay channels:
\begin{align}
\Gamma(D_s^+\to K^{*0} e^+\nu_{e})&=2.692_{-0.214}^{+0.182}\times 10^{-15}\,\rm GeV
\nonumber \\
\Gamma(D_s^+\to K^{*0} \mu^+\nu_{\mu})&=2.558_{-0.202}^{+0.173}\times 10^{-15}\,\rm GeV
\label{dGamma:numerical value}
\end{align}
furthermore, in Fig.~\ref{Fig:dGamma}, we present the decay widths for $D_s^+\to K^{*0}\ell^+\nu_{\ell}$ versus $q^2$. To facilitate comparison, experimental measurements from BESIII and predictions from other theoretical works are also included. From Fig.~\ref{Fig:dGamma}, the predictions given by the LCSR'25~\cite{Lin:2025cmn} has a significant deviation from the results calculated in this paper. All the other theoretical and experimental results are agree well with our calculations. Subsequently, by retrieving the lifetime of the final-state $D_s^+$ meson from PDG, the branching fraction of $D_s^+\to K^{*0}\ell^+\nu_{\ell}$ semileptonic decay is further calculated, and the results are listed in Table~\ref{Tab:III}. Also included in the table are other experimental and theoretical results, including PDG'24~\cite{ParticleDataGroup:2024cfk}, CLEO~\cite{CLEO:2009dyb,Hietala:2015jqa}, BESIII~\cite{BESIII:2018xre,BESIII:2026ceu}, CCQM~\cite{Ivanov:2019nqd,Soni:2018adu}, $\chi$UA'15~\cite{Sekihara:2015iha}, LFQM'17~\cite{Cheng:2017pcq}, HM$\chi$T'05~\cite{Fajfer:2005ug}, CQM'00~\cite{Melikhov:2000yu}, LCSR~\cite{Wu:2006rd,Lin:2025cmn}, CLFQM~\cite{Wang:2008ci,Zhang:2020dla,Yang:2025gfz}, and RQM'19~\cite{Faustov:2019mqr}. The analysis shows that the branching fraction of the semileptonic decay $D_s^+ \to K^{*0} \ell^+ \nu_\ell$ predicted in this work for both the electron and muon channels is in agreement within the error range, indicating that this decay process has good LFU. Compared with existing experimental results, our predicted results are highly consistent with the PDG~\cite{ParticleDataGroup:2024cfk} average values and also agree with the results from CLEO~\cite{CLEO:2009dyb,Hietala:2015jqa} and BESIII~\cite{BESIII:2018xre,BESIII:2026ceu}. On the theoretical side, our predictions are consistent with those of CCQM~\cite{Ivanov:2019nqd,Soni:2018adu}, $\chi$UA'15~\cite{Sekihara:2015iha}, LFQM'17~\cite{Cheng:2017pcq}, HM$\chi$T'05~\cite{Fajfer:2005ug}, CQM'00~\cite{Melikhov:2000yu},  LCSR~\cite{Wu:2006rd,Lin:2025cmn}, CLFQM'08~\cite{Wang:2008ci,Zhang:2020dla,Yang:2025gfz}, and RQM'19~\cite{Faustov:2019mqr} within the uncertainty ranges. Overall, the branching fraction results predicted in this work exhibit good self-consistency, and can provide a certain theoretical reference for future experimental re-measurements. Meanwhile, we also provided the correlation degree predictions for the branching fractions of two decay channels, $D_s^+ \to K^{*0}e^+\nu_e$ and $D_s^+ \to K^{*0}\mu^+\nu_\mu$, and presented them in Fig.~\ref{Fig:BRs}. From Fig.~\ref{Fig:BRs}, most theoretical predictions, fall inside or on the edges of the ellipses, while the central value of the BESIII'26 experiment is slightly higher but still compatible with the $2\sigma$ region. Overall, our results are in good agreement with existing experimental data and mainstream models, and the strong positive correlation between the electron and muon channels further supports LFU. In addition, we also calculated the ratio of branching fraction for two decay channels, $i,e.$, $\mathcal{R}_{\mu/e}^{K^{*0}}=\mathcal{B}(D_s^+\to K^{*0}\mu^+\nu_{\mu})/\mathcal{B}(D_s^+\to K^{*0}e^+\nu_{e})$. This ratio is an important probe of LFU in the semileptonic decays of the $D_s^+$ meson. Currently, the prediction range of this ratio by the SM is $[0.95, 0.99]$~\cite{BESIII:2026ceu,Ivanov:2019nqd,Cheng:2017pcq}. Using the obtained branching fraction, we can further calculate this ratio, and the result is $\mathcal{R}_{\mu/e}^{K^{*0}}=0.950_{-0.002}^{+0.004}$. The calculation results show that our prediction is consistent with the expectation of the SM.

Furthermore, we extract the CKM matrix element $|V_{cd}|$ within the framework of QCD LCSR, using the semileptonic branching fractions $\mathcal{B}(D_s^+\to K^{*0} e^+\nu_{e})$ and $\mathcal{B}(D_s^+\to K^{*0} \mu^+\nu_{\mu})$ measured by the BESIII'26~\cite{BESIII:2026ceu}. The results are as follows:
\begin{align}
|V_{cd}|_{(e \text{-Channel})}&=0.225_{-0.005}^{+0.005}
\nonumber \\
|V_{cd}|_{(\mu \text{-Channel})}&=0.227_{-0.004}^{+0.011}
\end{align}
meanwhile, we compared this results with those of other theories and experimental groups regarding the CKM matrix element $|V_{cd}|$, as shown in Fig.~\ref{Fig:CKM}. These results include PDG,24~\cite{ParticleDataGroup:2024cfk}, LQCD~\cite{FermilabLattice:2022gku,Lubicz:2017syv,Na:2011mc,Ke:2023qzc,Riggio:2017zwh}, FLAG'24~\cite{FlavourLatticeAveragingGroupFLAG:2024oxs}, BESIII~\cite{BESIII:2015tql,Ablikim:2020hsc,BESIII:2013iro}, CELO'09~\cite{CLEO:2009svp}, and BaBar'14~\cite{BaBar:2014xzf}. From Fig.~\ref{Fig:CKM} that our prediction results are consistent with the results of these experiments and theoretical studies within uncertainties.
\begin{figure}[t]
\begin{center}
\centering
\includegraphics[width=0.42\textwidth]{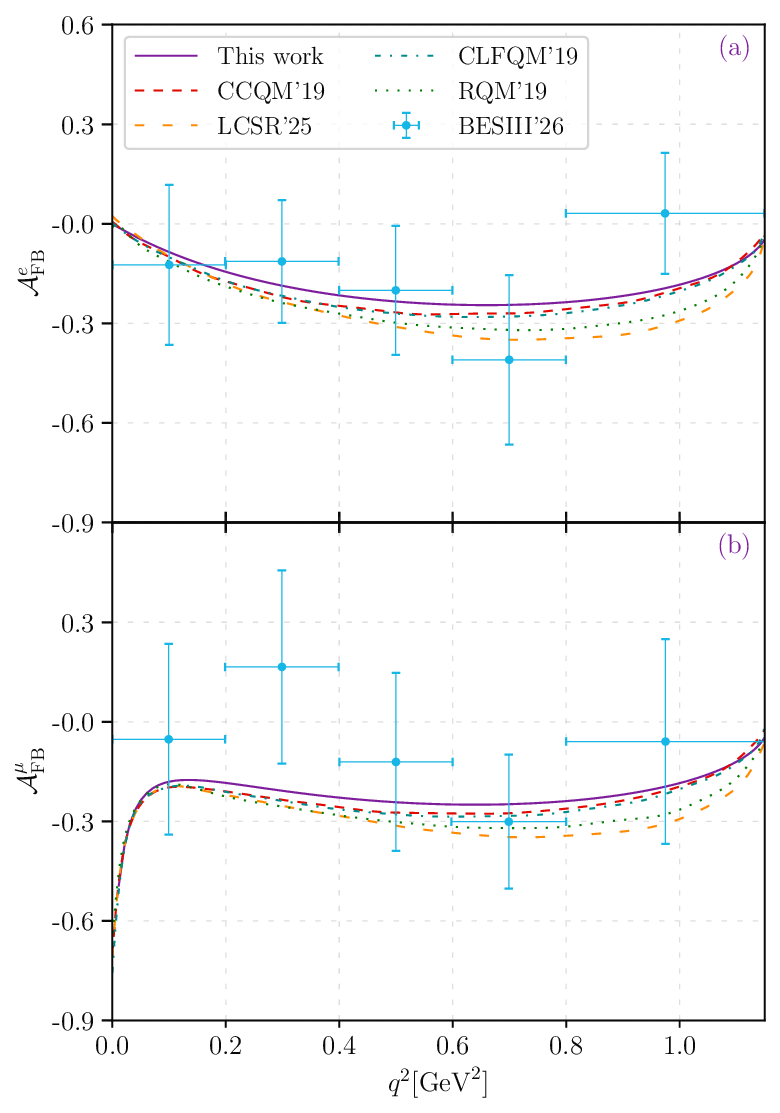}
\end{center}
\caption{Our predict the dependence of the forward-backward asymmetry $\mathcal{A}^{\ell}_{\rm FB}$ on the transferred momentum $q^2$ in the semileptonic decay $D_s^+ \to K^{*0}\ell^+\nu_{\ell}$ with $\ell = (e,\mu)$. Meanwhile, we also compare it with the experimental and theoretical results of BESIII'26~\cite{BESIII:2026ceu}, CCQM'19~\cite{Ivanov:2019nqd}, LCSR'25~\cite{Lin:2025cmn}, CLFQM'19~\cite{Zhang:2020dla}, and RQM'19~\cite{Faustov:2019mqr}.}
\label{Fig:AFB}
\end{figure}

Finally, based on the definition of forward-backward asymmetry $\mathcal{A}_{\rm FB}$, we derive its relationship with the transition momentum $q^2$. The corresponding results are present in Fig.~\ref{Fig:AFB}, alongside comparisons with the findings of BESIII'26~\cite{BESIII:2026ceu}, CCQM'19~\cite{Ivanov:2019nqd}, LCSR'25~\cite{Lin:2025cmn}, CLFQM'19~\cite{Zhang:2020dla}, and RQM'19~\cite{Faustov:2019mqr}. Furthermore, we calculate the average values of forward-backward asymmetry, obtaining $\langle \mathcal{A}_{\rm FB}^e \rangle = -0.212_{-0.003}^{+0.001}$ and $\langle \mathcal{A}_{\rm FB}^\mu \rangle = -0.243_{-0.003}^{+0.008}$, respectively. These results are consistent with the latest BESIII measurements, namely $\langle \mathcal{A}_{\rm FB}^e \rangle = -0.14 \pm 0.09_{\text{stat}} \pm 0.01_{\text{syst}}$ and $\langle \mathcal{A}_{\rm FB}^\mu \rangle = -0.12 \pm 0.10_{\text{stat}} \pm 0.01_{\text{syst}}$. Meanwhile, our theoretical predictions agree within $1\sigma$ with the theoretical predictions for the electron mode, ranging from $-0.22$ to $-0.26$, and for the muon mode, ranging from $-0.25$ to $-0.29$~\cite{Soni:2018adu,Ivanov:2019nqd,Faustov:2019mqr,ATLAS:2024vyj}. This also indicates that no significant evidence of LFU violation is observed in the semileptonic decays $D_s^+ \to K^{*0}\ell^+\nu_\ell$.

\section{Summary}\label{Sec:4}
In this work, we investigate the semileptonic decay $D_s^+\to K^{*0}$ within the QCD LCSRs framework. First, we construct the correlation function using usual currents to compute the corresponding TFFs for $D_s^+\to K^{*0}$. The longitudinal and transverse twist-2 LCDAs of the $K^{*0}$ meson is adopted as essential input parameters. The model parameters of these LCDAs can be fixed using the above constraints, from which their specific numerical values are determined, as listed in Table~\ref{Tab:I}. Fig.~\ref{Fig:DA} further shows the distributions of the $K^{*0}$ meson LCDAs $\phi^\lambda_{2;K^{*0}}(x,\mu_0)$. Subsequently, we calculate TFFs at the large recoil region and their ratios and compare them with existing experimental and theoretical results (see Tables~\ref{Tab:TFFs} and~\ref{Tab:RR}). Similarly, in Fig~\ref{Fig:ratios}, the correlation between TFFs and their ratios is also presented. Furthermore, the TFFs are extended to the entire physical $q^2$ region using a simplified $z(q^2,t_0)$ series expansion, thereby obtaining the behavior for TFFs over the whole region, as shown in Fig.~\ref{Fig:fp2}. Finally, by consulting the PDG and using the decay lifetime of the $D_s^+$ meson, we calculate the decay width integrals for the $D_s^+\to K^{*0}$ semileptonic decay, obtaining the results $\Gamma(D_s^+\to K^{*0} e^+\nu_{e})=2.692_{-0.214}^{+0.182}\times 10^{-15}\,\rm GeV$ and $\Gamma(D_s^+\to K^{*0} \mu^+\nu_{\mu})=2.558_{-0.202}^{+0.173}\times 10^{-15}\,\rm GeV$. Meanwhile, we calculate the differential decay width of the $D_s^+\to K^{*0}$ process, as shown in Fig.~\ref{Fig:dGamma}. The corresponding decay branching fractions are listed in Table~\ref{Tab:III}. Meanwhile, the correlation between the branching fractions of the two decay channels for this process is shown in Fig.~\ref{Fig:BRs}. From this, it can also be deduced that the ratio of branching fractions for the two decay channels is $\mathcal{R}_{\mu/e}^{K^{*0}}=0.950_{-0.002}^{+0.004}$. Subsequently, based on the branching fraction measured by the BESIII experiment, we extract the value of the CKM matrix element $|V_{cd}|$. For comparison, we also collect the results for $|V_{cd}|$ obtained from other theoretical and experimental studies, which are summarized in Fig.~\ref{Fig:CKM}. Finally, based on the definition of forward-backward asymmetry, we predict the dependence of $\mathcal{A}^{\ell}_{\rm FB}$ on the transferred momentum $q^2$ in this semileptonic decay process. The results are shown in Fig.~\ref{Fig:AFB}. In addition, the average values of $\mathcal{A}^{\ell}_{\rm FB}$ are given as $\langle \mathcal{A}_{\rm FB}^e \rangle = -0.212_{-0.03}^{+0.01}$ and $\langle \mathcal{A}_{\rm FB}^\mu \rangle = -0.243_{-0.003}^{+0.008}$, respectively. These results indicate that no significant violation of LFU is observed in the semileptonic decay $D_s^+ \to K^{*0} \ell^+ \nu_{\ell}$.
\\
\section{Acknowledgments}\label{Sec:5}
This work was supported by the National Natural Science Foundation of China under Grant No.12265010, the Project of Guizhou Provincial Department of Science and Technology under Grants No.MS[2025]219 and No.CXTD[2025]030, the Guizhou Provincial Scientific Research Project (Grant No.2025YJSKYJJ171) and the Scientific Research Project of
Guizhou Minzu University (Grant No.GZMUBS10).
\\

\appendix
\section{The auxiliary coefficients of the $D_s^+ \to K^{*0}$ TFFs}\label{sec:appendixA}
In the framework of LCSR, the auxiliary coefficients contained in the analytical expressions of the $D_s^+ \to K^{*0}$ TFFs are provided below, with their explicit forms listed one by one.
\begin{widetext}
\begin{eqnarray}
B_{A_1}&=&\frac{m_{K^{*0}} m_c}{f_{D_s^+} m_{D_s^+}^2(m_{D_s^+} + m_{K^{*0}})}\int_0^1 du e^{\left( {m_{{D_s^+}}^2 - s(u)} \right) / M^2}
\frac{ m_{K^{*0}} f_{K^{*0}}^\bot \cal C}{2u^2 m_{K^{*0}} ^2}\Theta(c(u,s_0)),
\nonumber\\
C_{A_1}&=&\frac{m_{K^{*0}} m_c}{f_{D_s^+} m_{D_s^+}^2(m_{D_s^+} + m_{K^{*0}})}\int_0^1 du e^{\left( {m_{{D_s^+}}^2 - s(u)} \right) / M^2}
\frac{m_{K^{*0}} f_{K^{*0}}^\bot}{2u} \Theta(c(u,s_0)),
\nonumber\\
D_{A_1}&=&\frac{m_{K^{*0}} m_c}{f_{D_s^+} m_{D_s^+}^2(m_{D_s^+} + m_{K^{*0}})}\int_0^1 du e^{\left( {m_{{D_s^+}}^2 - s(u)} \right) / M^2}
\frac{ m_c f_{K^{*0}}^\parallel}{u} \Theta \left(c \left( u,s_0 \right) \right),
\nonumber\\
E_{A_1}&=&\frac{m_{K^{*0}}^2 m_c f_{K^{*0}}^\bot}{f_{D_s^+} m_{D_s^+}^2(m_{D_s^+} + m_{K^{*0}})}\int_0^1 du e^{\left( {m_{{D_s^+}}^2 - s(u)} \right) / M^2}
\bigg[ \frac{m_c^2{\cal C}}{8u^4M^4} \widetilde{\widetilde\Theta}(c(u,s_0))\,+\, \frac{{\cal C}-2m_c^2}{8u^3M^2} \widetilde\Theta(c(u,s_0)) \frac{1}{8u^2}\Theta(c(u,s_0))\bigg],
\nonumber\\
F_{A_1}&=&\frac{m_{K^{*0}} m_c}{f_{D_s^+} m_{D_s^+}^2(m_{D_s^+} + m_{K^{*0}})}\int_0^1 du e^{\left( {m_{{D_s^+}}^2 - s(u)} \right) / M^2}
\frac{m_c m_{K^{*0}}^2 f_{K^{*0}}^\parallel }{u^2 M^2}\widetilde \Theta \left( c\left( u,s_0 \right) \right),
\nonumber\\
G_{A_1}&=&\frac{m_{K^{*0}}^2 m_c f_{K^{*0}}^\bot}{f_{D_s^+} m_{D_s^+}^2(m_{D_s^+} + m_{K^{*0}})}\int_0^1 du e^{\left( {m_{{D_s^+}}^2 - s(u)} \right) / M^2}
\bigg[\,\frac{\cal C}{u^3M^2}\widetilde\Theta(c(u,s_0))-\frac{1}{u^2} \Theta(c(u,s_0))\, \bigg],
\nonumber\\
H_{A_1}&=&\frac{m_{K^{*0}}^2 m_c f_{K^{*0}}^\bot}{f_{D_s^+} m_{D_s^+}^2(m_{D_s^+} + m_{K^{*0}})}\int_0^1 du e^{\left( {m_{{D_s^+}}^2 - s(u)} \right) / M^2}
\bigg[\,\frac{2m_c^2}{2u^2M^2}\widetilde \Theta(c(u,s_0))+\frac{1}{2u} \Theta(c(u,s_0))\,\bigg],
\nonumber\\
I_{A_1}&=&\int_0^1 dv \int_0^1 du \,\int_0^1 d {\mathcal D}\,e^{\left( {m_{B}^2 - s(u)} \right) / M^2}\frac{\widetilde\Theta(c(u,s_0))}{u^2\,M^2} \frac{m_c\,m_{K^{*0}}^2 f_{K^{*0}}^\bot} {12 f_{D_s^+} m_{D_s^+}^2\,\,( m_{D_s^+} + m_{K^{*0}})} \bigg(m_{D_s^+}^2-m_{K^{*0}}^2+2 u m_{K^{*0}}^2 \bigg),
\nonumber\\
J_{A_1}&=&\int_0^1 dv \int_0^1 du \,\int_0^1 d {\mathcal D}\,e^{\left( {m_{B}^2 - s(u)} \right) / M^2}\frac{\widetilde\Theta(c(u,s_0))}{u^2\,M^2} \frac{(1 - 2v) m_c\,m_{K^{*0}}^2 f_{K^{*0}}^\bot} {f_{D_s^+} m_{D_s^+}^2\,\,( m_{D_s^+} + m_{K^{*0}})} \bigg(m_{D_s^+}^2-m_{K^{*0}}^2+2 u m_{K^{*0}}^2 \bigg),
\nonumber\\
K_{A_1}&=&\int_0^1 dv \int_0^1 du \,\int_0^1 d {\mathcal D}\,e^{\left( {m_{B}^2 - s(u)} \right) / M^2}\frac{\widetilde\Theta(c(u,s_0))}{u^2\,M^2} \frac{2\,m_c\,m_{K^{*0}}^2 f_{K^{*0}}^\bot} {f_{D_s^+} m_{D_s^+}^2\,\,( m_{D_s^+} + m_{K^{*0}})} \bigg(m_{D_s^+}^2-m_{K^{*0}}^2+2 u m_{K^{*0}}^2 \bigg),
\nonumber\\
L_{A_1}&=&\int_0^1 dv \int_0^1 du \,\int_0^1 d {\mathcal D}\,e^{\left( {m_{B}^2 - s(u)} \right) / M^2}\frac{\widetilde\Theta(c(u,s_0))}{u^2\,M^2} \frac{(2 - 4v)\,m_c\,m_{K^{*0}}^2 f_{K^{*0}}^\bot} {f_{D_s^+} m_{D_s^+}^2\,\,( m_{D_s^+} + m_{K^{*0}})} \bigg(m_{D_s^+}^2-m_{K^{*0}}^2+2 u m_{K^{*0}}^2 \bigg),
\nonumber\\
M_{A_1}&=&\int_0^1 dv \int_0^1 du \,\int_0^1 d {\mathcal D}\,e^{\left( {m_{B}^2 - s(u)} \right) / M^2}\frac{\widetilde\Theta(c(u,s_0))}{u^2\,M^2} \frac{m_c^2\,m_{K^{*0}}^3 f_{K^{*0}}^\parallel} {6 f_{D_s^+} m_{D_s^+}^2\,\,( m_{D_s^+} + m_{K^{*0}})}.
\label{tra A1}
\\
B_{A_2}&=&\frac{m_{K^{*0}} m_c\left( m_{D_s^+} + m_{K^{*0}} \right)}{2f_{D_s^+} m_{D_s^+}^2}\int_0^1 du e^{\left( {m_{{D_s^+}}^2 - s(u)} \right) / M^2} \frac{m_{K^{*0}} f_{K^{*0}}^\bot}{u m_{K^{*0}} ^2}\Theta(c(u,s_0)),
\nonumber\\
C_{A_2}&=&\frac{m_{K^{*0}} m_c\left( m_{D_s^+} + m_{K^{*0}} \right)}{2f_{D_s^+} m_{D_s^+}^2}\int_0^1 du e^{\left( {m_{{D_s^+}}^2 - s(u)} \right) / M^2} \frac{m_c f_{K^{*0}}^\bot}{u M^2} \widetilde\Theta (c(u,s_0)),
\nonumber\\
D_{A_2}&=&\frac{m_{K^{*0}}^2 m_c f_{K^{*0}}^\bot \left( m_{D_s^+} + m_{K^{*0}} \right)}{8 f_{D_s^+} m_{D_s^+}^2}\int_0^1 du e^{\left( {m_{{D_s^+}}^2 - s(u)} \right) / M^2}
\bigg[\frac{m_c^2}{u^3M^4} \widetilde{\widetilde\Theta}(c(u,s_0))\,+\,\frac{1}{u^2M^2} \widetilde\Theta(c(u,s_0))\bigg],
\nonumber\\
E_{A_2}&=&\frac{m_{K^{*0}} m_c\left( m_{D_s^+} + m_{K^{*0}} \right)}{2f_{D_s^+} m_{D_s^+}^2}\int_0^1 du e^{\left( {m_{{D_s^+}}^2 - s(u)} \right) / M^2}
\frac{2 m_c f_{K^{*0}}^\parallel }{u^2 M^2}\widetilde \Theta \left( c\left( {u,{s_0}} \right) \right),
\nonumber\\
F_{A_2}&=&\frac{m_{K^{*0}} m_c\left( m_{D_s^+} + m_{K^{*0}} \right)}{2f_{D_s^+} m_{D_s^+}^2}\int_0^1 du e^{\left( {m_{{D_s^+}}^2 - s(u)} \right) / M^2}
\frac{m_{K^{*0}}^2 m_c^3 f_{K^{*0}}^\parallel}{2u^4 M^6}\,\widetilde {\widetilde{\widetilde \Theta }}\left( c\left( u,s_0 \right) \right),
\nonumber\\
G_{A_2}&=&\frac{m_{K^{*0}} m_c\left( m_{D_s^+} + m_{K^{*0}} \right)}{2f_{D_s^+} m_{D_s^+}^2}\int_0^1 du e^{\left( {m_{{D_s^+}}^2 - s(u)} \right) / M^2}
\frac{2 m_c m_{K^{*0}}^2 f_{K^{*0}}^\parallel }{u^2 M^4}\,\,\widetilde {\widetilde \Theta }\left( c\left( u,s_0 \right)\right),
\nonumber\\
H_{A_2}&=&\frac{m_{K^{*0}}^2 m_c f_{K^{*0}}^\bot \left( m_{D_s^+} + m_{K^{*0}} \right)}{f_{D_s^+} m_{D_s^+}^2}\int_0^1 du e^{\left( {m_{{D_s^+}}^2 - s(u)} \right) / M^2}
\bigg[\,\frac{{\cal C} \,-\, 2m_c^2}{u^3 M^4}\,\,\widetilde{\widetilde\Theta}(c(u,s_0)) - \frac{1}{u^2M^2}\,\widetilde\Theta (c(u,s_0))\bigg],
\nonumber\\
I_{A_2}&=&\frac{m_{K^{*0}} m_c\left( m_{D_s^+} + m_{K^{*0}} \right)}{2f_{D_s^+} m_{D_s^+}^2}\int_0^1 du e^{\left( {m_{{D_s^+}}^2 - s(u)} \right) / M^2}
\frac{m_{K^{*0}}\,f_{K^{*0}}^\bot}{u M^2} \widetilde \Theta (c(u,s_0)),
\nonumber\\
J_{A_2}&=&\int_0^1 dv \int_0^1 du \int_0^1 d {\mathcal D} e^{\left( {m_{{D_s^+}}^2 - s(u)} \right) / M^2} \frac{ m_c m_{K^{*0}}^2 f_{K^{*0}}^\bot} {12 f_{D_s^+} m_{D_s^+}^2} \frac{ m_{D_s^+} + m_{K^{*0}}}{u^2 M^2} \,\widetilde\Theta(c(u,s_0)),
\nonumber\\
K_{A_2}&=&\int_0^1 dv \int_0^1 du \int_0^1 d {\mathcal D} e^{\left( {m_{{D_s^+}}^2 - s(u)} \right) / M^2} \frac{(2v - 1)\, m_c m_{K^{*0}}^2 f_{K^{*0}}^\bot} {f_{D_s^+} m_{D_s^+}^2} \frac{ m_{D_s^+} + m_{K^{*0}}}{u^2 M^2} \,\widetilde\Theta(c(u,s_0)),
\nonumber\\
L_{A_2}&=&\int_0^1 dv \int_0^1 du \int_0^1 d {\mathcal D} e^{\left( {m_{{D_s^+}}^2 - s(u)} \right) / M^2} \frac{(4v - 2)\, m_c m_{K^{*0}}^2 f_{K^{*0}}^\bot} {f_{D_s^+} m_{D_s^+}^2} \frac{ m_{D_s^+} + m_{K^{*0}}}{u^2 M^2} \,\widetilde\Theta(c(u,s_0)),
\nonumber\\
M_{A_2}&=&\int_0^1 dv \int_0^1 du \int_0^1 d {\mathcal D} e^{\left( {m_{{D_s^+}}^2 - s(u)} \right) / M^2} \frac{2\, m_c m_{K^{*0}}^2 f_{K^{*0}}^\bot} {f_{D_s^+} m_{D_s^+}^2} \frac{ m_{D_s^+} + m_{K^{*0}}}{u^2 M^2} \,\widetilde\Theta(c(u,s_0)).
\label{tra A2}
\\
B_{V}&=&\frac{m_c \left( m_{D_s^+} + m_{K^{*0}} \right)}{2f_{D_s^+} m_{D_s^+}^2}\int_0^1 du e^{\left( {m_{{D_s^+}}^2 - s(u)} \right) / M^2}
f_{K^{*0}}^\bot\,\Theta(c(u,s_0)),
\nonumber\\
C_{V}&=&\frac{m_c \left( m_{D_s^+} + m_{K^{*0}} \right)}{2f_{D_s^+} m_{D_s^+}^2}\int_0^1 du e^{\left( {m_{{D_s^+}}^2 - s(u)} \right) / M^2}
\frac{m_{K^{*0}} m_c f_{K^{*0}}^\parallel}{2u^2 M^2}\widetilde \Theta \left( c\left( u,s_0 \right) \right),
\nonumber\\
D_{V}&=&\frac{m_{K^{*0}}^2 m_c f_{K^{*0}}^\bot \left( m_{D_s^+} + m_{K^{*0}} \right)}{8 f_{D_s^+} m_{D_s^+}^2}\int_0^1 du e^{\left( {m_{{D_s^+}}^2 - s(u)} \right) / M^2} \bigg[\,\,\frac{m_b^2}{u^2M^4}\widetilde {\widetilde\Theta}(c(u,s_0))+\frac{1}{uM^2}\widetilde\Theta(c(u,s_0))\,\,\bigg],
\nonumber\\
E_{V}&=&\int_0^1 dv \int_0^1 du \int_0^1 d {\mathcal D} e^{\left( {m_{{D_s^+}}^2 - s(u)} \right) / M^2}\,\frac{(2v - 1)\, \widetilde\Theta(c(u,s_0))\, m_{K^{*0}}^2\,f_{K^{*0}}^\bot}{6\, u^2 M^2},
\nonumber\\
F_{V}&=&\int_0^1 dv \int_0^1 du \int_0^1 d {\mathcal D} e^{\left( {m_{{D_s^+}}^2 - s(u)} \right) / M^2}\,\frac{2\, \widetilde\Theta(c(u,s_0))\,m_{K^{*0}}^2\,f_{K^{*0}}^\bot}{u^2 M^2},
\nonumber\\
G_{V}&=&\int_0^1 dv \int_0^1 du \int_0^1 d {\mathcal D} e^{\left( {m_{{D_s^+}}^2 - s(u)} \right) / M^2}\,\frac{4(v - 1)\, \widetilde\Theta(c(u,s_0))\,m_{K^{*0}}^2\,f_{K^{*0}}^\bot}{u^2 M^2}.
\label{tra V}
\\
B_{A_{30}}&=&- \frac{m_c q^2}{2f_{D_s^+} m_{D_s^+}^2}\int_0^1 du e^{\left( {m_{{D_s^+}}^2 - s(u)} \right) / M^2}
\frac{f_{K^*}^\bot}{2u m_{K^*}}\,\Theta (c(u,s_0)),
\nonumber\\
C_{A_{30}}&=&\frac{m_c q^2}{2f_{D_s^+} m_{D_s^+}^2}\int_0^1 du e^{\left( {m_{{D_s^+}}^2 - s(u)} \right) / M^2}
\frac{(2-u)\,m_{K^{*0}}f_{K^{*0}}^\bot}{2u^2M^2}\,\widetilde \Theta (c(u,s_0)),
\nonumber\\
D_{A_{30}}&=&\frac{m_c q^2}{2f_{D_s^+} m_{D_s^+}^2}\int_0^1 du e^{\left( {m_{{D_s^+}}^2 - s(u)} \right) / M^2} \frac{m_{K^{*0}}f_{K^{*0}}^\bot}{8}\bigg[\frac{m_c^2}{u^3M^4} \widetilde{\widetilde\Theta}(c(u,s_0))+ \frac{1}{u^2M^2} \widetilde\Theta(c(u,s_0))\bigg],
\nonumber\\
E_{A_{30}}&=&\frac{m_c q^2}{2f_{D_s^+} m_{D_s^+}^2}\int_0^1 du e^{\left( {m_{{D_s^+}}^2 - s(u)} \right) / M^2}
\frac{m_c f_{K^{*0}}^\|}{m_{K^{*0}}u^2M^2}\widetilde \Theta (c(u,s_0)),
\nonumber\\
F_{A_{30}}&=&\frac{m_c q^2}{2f_{D_s^+} m_{D_s^+}^2}\int_0^1 du e^{\left( {m_{{D_s^+}}^2 - s(u)} \right) / M^2}
\frac{m_{K^{*0}} m_c^3 f_{K^{*0}}^\|}{u^4 M^6}\,\widetilde {\widetilde {\widetilde \Theta }}(c(u,s_0)),
\nonumber\\
G_{A_{30}}&=&\frac{m_c q^2}{2f_{D_s^+} m_{D_s^+}^2}\int_0^1 du e^{\left( {m_{{D_s^+}}^2 - s(u)} \right) / M^2}
\frac{m_cm_{K^{*0}}f_{K^{*0}}^\|(2 - u)}{u^3 M^4}\widetilde {\widetilde \Theta }(c(u,s_0)),
\nonumber\\
H_{A_{30}}&=&\frac{m_{K^{*0}} m_c q^2 f_{K^{*0}}^\bot}{2f_{D_s^+} m_{D_s^+}^2}\int_0^1 du e^{\left( {m_{{D_s^+}}^2 - s(u)} \right) / M^2}
\bigg[(4 - 2u)\bigg(\frac{\cal C}{2u^4 M^4} \widetilde{\widetilde\Theta}(c(u,s_0)) - \frac{1}{u^3 M^2}\widetilde\Theta(c(u,s_0))\bigg)
\nonumber\\
&& -\, \bigg(\frac{2m_c^2}{u^3 M^4}\widetilde{\widetilde\Theta}(c(u,s_0)) + \frac{1}{u^2M^2} \widetilde\Theta(c(u,s_0))\bigg)\bigg],
\nonumber\\
I_{A_{30}}&=&\frac{m_c q^2}{2f_{D_s^+} m_{D_s^+}^2}\int_0^1 du e^{\left( {m_{{D_s^+}}^2 - s(u)} \right) / M^2}
\frac{m_{K^{*0}}(2-u)}{2u^2M^2} f_{K^{*0}}^\bot\widetilde\Theta(c(u,s_0)),
\nonumber\\
J_{A_{30}}&=&\int_0^1 dv \int_0^1 du \int_0^1 d {\mathcal D}  e^{\left( {m_{{D_s^+}}^2 - s(u)} \right) / M^2}\frac{(2v -1) m_c q^2 f_{K^{*0}}^\bot} {12 f_{D_s^+} m_{D_s^+}^2} \frac{\widetilde\Theta(c(u,s_0))}{u^2 M^2},
\nonumber\\
K_{A_{30}}&=&\int_0^1 dv \int_0^1 du \int_0^1 d {\mathcal D}  e^{\left( {m_{{D_s^+}}^2 - s(u)} \right) / M^2}\frac{(4v-2) m_c q^2 f_{K^{*0}}^\bot} {12 f_{D_s^+} m_{D_s^+}^2} \frac{\widetilde\Theta(c(u,s_0))}{u^2 M^2},
\nonumber\\
L_{A_{30}}&=&\int_0^1 dv \int_0^1 du \int_0^1 d {\mathcal D}  e^{\left( {m_{{D_s^+}}^2 - s(u)} \right) / M^2}\frac{ m_c q^2 f_{K^{*0}}^\bot} {f_{D_s^+} m_{D_s^+}^2} \frac{\widetilde\Theta(c(u,s_0))}{u^2 M^2},
\nonumber\\
M_{A_{30}}&=&\int_0^1 dv \int_0^1 du \int_0^1 d {\mathcal D}  e^{\left( {m_{{D_s^+}}^2 - s(u)} \right) / M^2}\frac{ m_c q^2 f_{K^{*0}}^\bot} {24 f_{D_s^+} m_{D_s^+}^2} \frac{\widetilde\Theta(c(u,s_0))}{u^2 M^2}.
\label{tra A3-A0}
\end{eqnarray}
\end{widetext}
where the $\int d{\cal D} = \int d\alpha_1\,d\alpha_2 d\alpha_3 \,\delta\,(1 - \sum_{i=1}^{3} \alpha_i)$, step function $\Theta(c(u,s_0))$ and surface terms $\delta(c(u_0,s_0))$ and $\Delta(c(u_0,s_0))$ can be found in Ref.~\cite{Fu:2014uea,Cheng:2017bzz}. Meanwhile, the following auxiliary quantities:
${\mathcal H} = q^2/(m_{D_s^+}^2 - m_{K^{*0}}^2)$,
${\mathcal E} = m_c^2 - u^2 m_{K^{*0}}^2 + q^2$,
${\mathcal C} = m_c^2 + u^2 m_{K^{*0}}^2 - q^2$,
$\mathcal Q = m_{D_s^+}^2 - m_{K^{*0}}^2 - q^2$,
$\mathcal F = m_c^2 - u^2 m_{K^{*0}}^2 - q^2$,
$c(\varrho,s_0) = \varrho s_0 - m_c^2 + \bar \varrho q^2 - \varrho \bar \varrho m_{K^{*0}}^2$,
and
$s(\varrho) = \left[ m_c^2 - \bar \varrho (q^2 - \varrho m_{K^{*0}}^2) \right] / \varrho$,
where $\varrho = u$ and $\bar \varrho = 1 - \varrho$. These quantities are essential for the subsequent analysis. Furthermore, the reduced functions $I_L(u)$, $H_3(u)$, $A_{K^{*0}}(u)$, $B_{K^{*0}}(u)$ and $C_{K^{*0}}(u)$ in the system of Eqs.~\eqref{tra A1}-\eqref{tra A3-A0} can be determined by referring to Refs.~\cite{Fu:2014uea,Cheng:2017bzz}. Their specific definitions are as follows
\begin{eqnarray}
I_L(u)&=& \int_0^u dv \int_0^v dw \bigg[\phi_{3;{K^{*0}}}^\|(w) -\frac{1}{2} \phi_{2;{K^{*0}}}^\bot(w)-\frac{1}{2} \psi_{4;{K^{*0}}}^\bot(w)\bigg],
\nonumber\\
H_3(u)&=& \int_0^u dv \bigg[\psi_{4;{K^*}}^{\bot}(v)-\phi_{2;{K^{*0}}}^\bot(v)\bigg],
\nonumber\\
A_{K^{*0}}(u) &=&\int_0^u dv\bigg[ \phi _{2;K^{*0}}^\| (v) - \phi _{3;K^{*0}}^ \bot (v)\bigg],
\nonumber\\
B_{K^{*0}}(u) &=&\int_0^u dv \phi_{4;K^{*0}}^\| (v),
\nonumber\\
C_{K^{*0}}(u) &\!=&\!\!\int_0^u\!dv\!\!\!\int_0^v\!dw\bigg[\psi _{4;K^{*0}}^\|(w)\!+\!\phi_{2;K^{*0}}^\|(w)\!-\! 2 \phi_{3;K^{*0}}^\bot(w)\bigg].
\label{TFFs: RF}
\end{eqnarray}
In Eq~\eqref{TFFs: RF}, the $K^{*0}$ meson twist-3 LCDA, we adopt the WW approximation~\cite{Ball:1997rj} to express it in terms of the twist-2 LCDA. The twist-4 contributions are treated following Ref.~\cite{Ball:2007zt}. There are certain approaches that can help to limit the sources of uncertainty, thereby enhancing the reliability of the LCSR theoretical predictions.

\end{document}